\documentstyle[aps,psfig,floats]{revtex}
\def\bra#1{\left< #1 \right|}
\def\ket#1{\left| #1 \right>}
\def\layermath#1{
  \mathchoice
  {\hbox{$\displaystyle      #1 \mathsurround=0pt$}}
  {\hbox{$\textstyle         #1 \mathsurround=0pt$}}
  {\hbox{$\scriptstyle       #1 \mathsurround=0pt$}}
  {\hbox{$\scriptscriptstyle #1 \mathsurround=0pt$}}
}
\def\threedelims#1#2#3#4#5{\mathinner{\layermath{
  \mathopen{\left#1 #2 \vphantom{#4} \right#3}
  \mathclose{\left. #4 \vphantom{#2} \right#5}
  \nulldelimiterspace=0pt
}}}
\def\braket#1#2{\threedelims<{#1}|{#2}>}
\def\matrixel#1#2#3{\mathinner{
  \mathopen{\left< #1 \vphantom{#2#3} \right|}
  #2
  \mathclose{\left| #3 \vphantom{#1#2} \right>}
}}
\def\ketbra#1#2{
  \mathord{\left| #1 \vphantom{#2} \right>}
  \mathord{\left< #2 \vphantom{#1} \right|}
}
\def\avg#1{\left< #1 \right>}
\def\abs#1{\left| #1 \right|}

\def\Real{\mathop{\rm Re}\nolimits}
\def\Imag{\mathop{\rm Im}\nolimits}

\def\recip#1{\frac{1}{#1}}

\def\bvec#1{{\bf #1}}
\def\vhat#1{\hat{\bvec{#1}}}

\def\maybeminus#1{\if-#1-\fi}
\def\p#1#2{\maybeminus{#1}q,\maybeminus{#2}p}
\def\pp#1#2{\braket{\p++}{\p{#1}{#2}}}
\def\savg#1{\langle #1 \rangle}
\def\pqavg#1{\matrixel{\p++}{#1}{\p++}}
\def\MCl{M_{\text{Cl}}}
\def\ketCl#1{\left| #1 \right)}
\def\braketCl#1#2{\threedelims({#1}|{#2})}
\def\matrixelCl#1#2#3{\mathinner{
  \mathopen{\left( #1 \vphantom{#2#3} \right|}
  #2
  \mathclose{\left| #3 \vphantom{#1#2} \right)}
}}

\begin{document}

\title{Localization of Eigenfunctions in the Stadium Billiard}
\author{W. E. Bies\cite{bies-email}}
\address{Department of Physics, \\
  Harvard University, Cambridge, MA 02138}
\author{L. Kaplan}
\address{Department of Physics and Society of Fellows, \\
  Harvard University, Cambridge, MA 02138}
\author{M. R. Haggerty}
\address{Department of Physics, \\
  Harvard University, Cambridge, MA 02138}
\author{E. J. Heller}
\address{Department of Physics and Department of Chemistry and
  Chemical Biology, \\
  Harvard University, Cambridge, MA 02138}

\maketitle

\begin{abstract}%
    We present a systematic survey of scarring and symmetry effects in
    the stadium billiard.  The localization of individual
    eigenfunctions in Husimi phase space is studied first, and it is
    demonstrated that on average there is more localization than can
    be accounted for on the basis of random-matrix theory, even after
    removal of bouncing-ball states and visible scars.  A major point
    of the paper is that symmetry considerations, including parity
    and time-reversal symmetries, enter to influence
    the total amount of localization.  The properties of the local
    density of states spectrum are also investigated, as a function of
    phase space location.  Aside from the bouncing-ball region of
    phase space, excess localization of the spectrum is found on short
    periodic orbits and along certain symmetry-related lines; the
    origin of all these sources of localization is discussed
    quantitatively and comparison is made with analytical predictions.
    Scarring is observed to be present in all the energy ranges
    considered.  In light of these results the excess localization in
    individual eigenstates is interpreted as being primarily due to
    symmetry effects; another source of excess localization, scarring
    by multiple unstable periodic orbits, is smaller by a factor of
    $\sqrt{\hbar}$.
\end{abstract}

\section{Introduction}

According to scar theory, the quantum eigenfunctions of a classically
chaotic dynamical system do not always look locally like random
superpositions of plane waves with fixed energy, as predicted by Berry
\cite{berry}; instead, many eigenfunctions display a concentration of
amplitude around short unstable periodic orbits greater than that
expected on the basis of random matrix theory fluctuations.  The first
examples of scarring in the stadium billiard were presented by
Heller\cite{H1}, who also gave a semiclassical theory of
scarring based on dynamics in the linearizable region
around the periodic orbit (see also unpublished numerical work by
McDonald\cite{mcdonald}). Recent developments\cite{KH2} have
extended the theory of scars to the non-linearizable regime, to
include the effects of homoclinic recurrences at long times.  Scarring
is then seen to be a weak localization phenomenon coming from the
short-time correlations associated with an unstable periodic orbit; in
the energy domain, this means that a wavepacket centered on the orbit
may have non-random overlaps with the eigenstates of the system.
Quantitative measures of the strength of scarring have been developed and
tested numerically\cite{KH2,scarmeasures}.

In this paper we ask whether the localization caused by scarring on
not-too-unstable (i.e., short) periodic orbits, and by atypical
regions such as the ``bouncing ball''~\cite{Baecker,Tanner,otherbb}
modes is adequate to predict measures of localization in eigenstates
of chaotic systems.  Quantitative numerical confirmation of scar
theory\cite{KH2,scarmeasures} was limited to discrete-time maps, and
it would be desirable to study scarring quantitatively in the context
of more realistic and experimentally realizable systems, such as the
stadium (Bunimovich) billiard.  In what follows we present a
systematic study of eigenstate localization in the stadium billiard
(see Ref.~\cite{stadium} for some other recent analyses of this
system).  In Section~\ref{method}, we first present the numerical
method used to find the eigenstates.  Then, in
Section~\ref{sec:locprops}, we study the localization properties of
individual eigenstates, followed by an investigation in
Section~\ref{locdens} of the localization properties of the local
density of states.  We find evidence of scarring as a ubiquitous
phenomenon, in all the energy ranges considered.  We examine
quantitative predictions of scarring strengths based on the classical
structure around the unstable periodic orbits~\cite{KH2}.  However,
scar theory is not sufficient to explain all the observed
localization.  Symmetry effects, including the imprints of both time-reversal
and parity symmetries, are ultimately found to dominate the
excess wavefunction localization in the stadium.

\section{Method}
\label{method}

We study eigenstates of the time-independent Schr\"odinger equation in
an infinite 2-dimensional potential well in the shape of a stadium,
taken here to consist of a square of side 2 with semicircular endcaps
of radius 1.  We use the plane wave method~\cite{planewavemethod} to
find states with even parity with respect to reflection about each of
the two symmetry axes of the stadium.  It is assumed that eigenstates
at energy $E$ can be approximated as a superposition of plane waves at
that energy; i.e., with $|{\bf k}|^2 = E$ (we use $\hbar =
1$ and $m = 1/2$ here and throughout).
Therefore we use as a basis plane waves with
even-even parity, namely $\cos(k x \cos\theta_j) \cos(k y
\sin\theta_j)$, with $N$ values of $\theta_j$ chosen uniformly between
$0$ and $\pi/2$.  Then the coefficients of these $N$ plane waves are
determined by minimizing the squared value of the wavefunction at $M
\ge N$ equally spaced points on the boundary.  The wavefunction is set
to an arbitrary non-zero value at a point in the interior in order not
to have a singular system of equations, and $M$ is kept slightly
larger than $N$ so that the system of equations is overdetermined.  In
practice $M$ must be chosen so that there are several points per
wavelength.

Since the eigenvalues are not known {\it a priori}, we search for them
using the `tension' method: at each energy $E$ we solve the linear
system for the coefficients of the plane waves by singular-value
decomposition; then, we compute the integral of $|\psi|^2$ along the
boundary.  If $E$ is an eigenvalue this integral should vanish, to
within numerical precision.  In practice one finds sharp minima in the
tension as a function of $k$ that are identified with the eigenvalues.
Since it is not known in advance how many plane waves are needed to
give an accurate representation of the wavefunction at a given energy,
we repeat the search for minima with increasing numbers of plane waves
$N$ and boundary points $M$ until the results of subsequent iterations
agree; we find, for instance, that for $k$ around 100, $N = 300$ and
$M = 310$ are necessary.  This corresponds to eight points per
wavelength along the part of the boundary lying in the first quadrant.
The number of plane waves needed scales linearly with $k$.

The computed eigenvalues and eigenfunctions satisfy three diagnostic
tests: (1) the total density of states agrees well with the Weyl area
rule; the deviation is never more than about 5 states out of 5000, or 
0.1\%
(the bulk of this deviation
is due to a small periodic modulation in the level staircase function,
which has the right period to be attributable to the bouncing-ball states
\cite{Tanner};
after taking out this modulation
we find that the total number of
possible missing or extra states in the region {$100 < k < 200$} is
not more than one or two); 
(2) the histogram of level spacings is consistent with the
Wigner surmise for random-matrix statistics;
(3) the overlaps between eigenstates are found to be less than 1\%.  A
problem with our procedure is that it produces many shallow tension
minima, by which we mean that the contrast between the tension at a
local minimum and the background tension is less than a factor of 100.
The frequency of such shallow minima grows with increasing energy.
Moreover, these poor minima are not improved by increasing the
parameters $M$ and $N$ because doing so makes the matrix more
ill-conditioned.  This suggests that there is an intrinsic shortcoming
in the ability of our basis to represent eigenstates; to have a
perfectly accurate representation one would presumably need to include
evanescent waves of total energy $E$ as well.  Although it has been
shown by Berry \cite{berry2} that an evanescent wave may be
represented by a singular superposition of plane waves, this
representation is numerically ill behaved, giving poor convergence
properties.

In spite of this difficulty, essentially all of the questionable
states must correspond to true eigenstates, in order for the counting
of states to come out right.  Also, many of the shallow minima we
assume correspond to eigenstates occur in close conjunction with other
minima, and yet the numerically computed histogram of level spacings
decreases linearly toward zero at small spacings, in good agreement
with the theory.

Another difficulty with the present method is that the computation
time per eigenstate scales as $k^3$, making the collection of
extensive eigenstate statistics above $k=300$ almost prohibitively
expensive.  The plane-wave method described here has been improved
upon by Li and Hu~\cite{li}, who expand the derivative of the tension
analytically in $k$ in order to predict the position of the next
minimum.  Li and Hu's approach improves the speed of the plane-wave
method by a factor of about five, but does not change the way the
computation time scales with $k$.  Thus, only slightly higher values
of $k$ could be reached for a given computation time, and we have not
needed to employ this approach here.  Another approach to the stadium
billiard has been outlined by Saraceno's group \cite{saraceno}.  Their
method, suitable for very high energies, obtains all the eigenvalues
in a narrow range of $k$ simultaneously by solving a generalized
eigenvalue problem, rather than searching for them step by step.  The
energies to be studied in this paper, however, are not so high as to
necessitate the use of such an algorithm.

Lists of eigenstates of the even-even symmetry class in the three
energy ranges of $k=20$ to $30$, $k=100$ to $150$ and $k=200$ to $225$
were generated according to the method outlined above.  Some examples
are given in Fig.~\ref{fig:wavefunctions} of eigenfunctions
$\Psi_n(x,y)$ in coordinate space.

\subsection{Surface wavefunctions}

Classically, one typically uses the boundary of the billiard as a
Poincar\'e surface of section.  The variables parameterizing the
surface are $q$, the arclength along the boundary, and its conjugate
momentum $p$, which is the component of momentum parallel to the wall
(both positive in the clockwise sense).  $q$ is taken modulo the
length $L \equiv 4 + 2\pi$ of the perimeter, while $p$ is limited by
energy considerations to $-k \leq p \leq k$.  Then the classical
billiard dynamics is reduced to a nonlinear one-bounce map, $(q',p')^T
= \MCl (q,p)^T$.

A natural way of reducing the quantum problem from two to one
dimension is to characterize the Dirichlet
eigenfunctions $\Psi_n(x,y)$, which
are defined on the interior of the stadium, by their normal
derivatives $\phi_n(q)$ on the boundary:
\begin{equation}
    \braket{q}{n} \equiv \phi_n(q)
    \propto \vhat{n} \cdot \nabla \Psi_n(x,y)
    \label{eq:phi-def}
    .
\end{equation}
We will study the properties of these {\em surface wavefunctions}.
The wavefunctions are normalized according to the convention
\begin{equation}
\int \int dx dy |\Psi_n(x,y)|^2 \,,
\label{area-norm}
\end{equation}
which is equivalent \cite{Boasman} to the following condition on the
surface wavefunctions:
\begin{equation}
{1 \over {2k^2}} \int dq (\vhat{n}(q) \cdot {\bf r}(q)) |\phi_n(q)|^2 = 1\,,
\end{equation}
where {$\vhat{n}(q)$} is the unit normal at the position {$q$} along
the boundary, and {${\bf r}(q)$} is the displacement vector from the
center of the stadium to the position {$q$}.

The Husimi representation of a surface wavefunction $\ket{n}$ is given
by its projection $|\braket{q_0,p_0}{n}|^2$ onto test state Gaussians of
the form
\begin{equation}
    \braket{q}{q_0,p_0} = \frac{1}{\sigma^{1/2} \pi^{1/4}}
    \exp\left[-\frac{(q - q_0)^2}{2 \sigma^2} + i p_0 (q - q_0) \right]
    \label{eq:gaussian}
    .
\end{equation}
The Gaussian is centered at the point $(q_0,p_0)$ in the boundary
phase space.  The parameter $\sigma$ controls the aspect ratio of the
Gaussian: its width in position is $\sigma/\sqrt 2$ while its width in
momentum space is $1/(\sigma \sqrt{2})$.
To maintain a given aspect
ratio of the Gaussian in phase-space, $\sigma$ must scale as
$k^{-1/2}$.  Here, we choose $\sigma = \left[ (2 + \pi) / k
\right]^{1/2}$, which makes the aspect ratio unity in units where the
full phase space on the billiard boundary is taken to be a square.
Note that the wavefunctions are real in coordinate space but complex
in phase space.  Some example Husimi plots are given in
Fig.~\ref{fig:wavefunctions}.

\section{IPR statistics for individual eigenstates}
\label{sec:locprops}

We start by investigating the localization properties of individual
eigenstates.  As a measure of the degree of localization of the
eigenstates in phase space, we use the mean squared value of the
intensity, also known as the inverse participation ratio (IPR)
\begin{equation}
    {\rm IPR}_n = {{{1 \over J}\sum_{j=1}^J
        \abs{\braket{q_j,p_j}{n}}^4}
      \over {\left ({1 \over J}\sum_{j=1}^J
        \abs{\braket{q_j,p_j}{n}}^2\right)^2}} \,,
    \label{iprn}
\end{equation}
where $q_j$ and $p_j$ range over the entire phase space as $j$ varies
on a sufficiently fine grid, which we take to be $200 \times 200$.
The range of momenta $p_j$
covered is from $-k$ to $+k$ for a wavefunction with energy
$k^2$; outside this
classically allowed region there is almost no wavefunction amplitude.
The IPR measures how much fluctuation across phase space there is in
the eigenfunction.  If $\abs{\braket{q_j,p_j}{n}}^2$ were completely
uniform over phase space, the IPR would equal 1.  On the other hand,
if all the intensity were concentrated entirely at one point $j$ and was
zero elsewhere, the IPR would reach its maximal value $J$.  [Of
course, the uncertainty principle prevents the IPR from ever becoming
greater than the size of phase space in units of a Planck cell, i.e.,
$O(kL/2\pi)$.]  Random-matrix theory (RMT) predicts a Porter-Thomas
distribution of wavefunction intensities, which yields an IPR of 2
(for complex wavefunctions), already larger than the naive classical
expectation of 1.

A subtle point is that the wavefunction intensity
$\abs{\braket{q_j,p_j}{n}}^2$, even when averaged
over wavefunctions $|n\rangle$, is not uniform over the boundary phase 
space, being instead a non-trivial function of $p_j$
(see Eq.~(\ref{eq:mean-intensity}) below). Therefore, even in the absence
of quantum fluctuations, if the intensities 
$\abs{\braket{q_j,p_j}{n}}^2$
for a wavefunction
$|n \rangle$ were given simply by the mean value $\sqrt{1-(p/k)^2}$ as
in Eq.~\ref{eq:mean-intensity}, we would obtain an IPR of
\begin{equation}
\left\langle 1-(p/k)^2 \right\rangle / 
\left\langle \sqrt{1-(p/k)^2} \right\rangle ^2
\approx 1.08 \,.
\end{equation}
The averages $\langle \cdots \rangle$ here are of course taken over the
classically allowed range of $p$: $-k < p <k$.
If the intensity fluctuations around the smooth value
$\sqrt{1-(p/k)^2}$ were instead given by a Porter-Thomas distribution, we
would obtain an IPR of $1.08 \times 2=2.16$ (since the smooth behavior
and the Porter-Thomas fluctuations are independent of one another, the IPR
contributions simply multiply).

In Table~\ref{table:IPRs} the IPRs of the first $36$ eigenfunctions in
each energy range are tabulated. 
\begin{table}
    \begin{tabular}{|@{\qquad}dd@{\quad}|%
                     @{\qquad}dd@{\quad}|%
                     @{\qquad}dd@{\quad}|}
        \multicolumn{1}{|@{\qquad}c}{$k$} &
            \multicolumn{1}{c@{\quad}|@{\qquad}}{IPR} &
        \multicolumn{1}{c}{$k$} &
            \multicolumn{1}{c@{\quad}|@{\qquad}}{IPR} &
        \multicolumn{1}{c}{$k$} &
            \multicolumn{1}{c@{\quad}|}{IPR} \\
        \hline
        20.082 & 2.03 & 100.031 & 2.24 & 200.015 & 2.81 \\
        20.237 & 4.$43^\dagger$ & 100.047 & 2.88 & 200.047 & 3.24 \\
        20.439 & 2.47 & 100.086 & 2.$73^*$ & 200.054 & 2.41 \\
        20.457 & 13.$36^*$ & 100.099 & 7.$61^*$ & 200.076 & 2.77 \\
        20.728 & 5.$02^*$ & 100.133 & 2.56 & 200.099 & 2.76 \\
        20.825 & 1.47 & 100.224 & 3.$51^\dagger$ & 200.108 & 17.$91^*$ \\
        21.039 & 1.88 & 100.226 & 2.74 & 200.113 & 2.70 \\
        21.223 & 2.$84^\dagger$ & 100.273 & 2.02 & 200.116 & 2.94 \\
        21.285 & 2.57 & 100.285 & 2.05 & 200.123 & 3.18 \\
        21.322 & 2.11 & 100.342 & 2.02 & 200.151 & 2.54 \\
        21.671 & 5.$35^\dagger$ & 100.363 & 3.$54^\dagger$ & 200.190 & 2.68 \\
        21.717 & 2.61 & 100.402 & 2.64 & 200.195 & 3.11 \\
        21.948 & 3.12 & 100.416 & 2.38 & 200.212 & 2.79 \\
        22.119 & 1.71 & 100.496 & 2.20 & 200.255 & 2.66 \\
        22.284 & 1.69 & 100.535 & 7.$55^*$ & 200.266 & 2.45 \\
        22.504 & 3.12 & 100.552 & 2.42 & 200.296 & 2.21 \\
        22.674 & 3.27 & 100.613 & 2.57 & 200.317 & 2.86 \\
        22.839 & 2.81 & 100.628 & 2.62 & 200.328 & 2.$96^\dagger$ \\
        23.029 & 2.41 & 100.678 & 1.93 & 200.334 & 2.$77^*$ \\
        23.108 & 3.$39^\dagger$ & 100.712 & 2.43 & 200.351 & 2.56 \\
        23.361 & 2.44 & 100.714 & 2.28 & 200.370 & 2.15 \\
        23.503 & 1.82 & 100.787 & 2.35 & 200.386 & 2.12 \\
        23.594 & 15.$79^*$ & 100.830 & 3.94 & 200.407 & 3.12 \\
        23.774 & 2.10 & 100.846 & 3.53 & 200.441 & 2.30 \\
        23.821 & 3.08 & 100.854 & 2.68 & 200.480 & 2.37 \\
        23.904 & 2.63 & 100.933 & 4.$95^*$ & 200.492 & 2.50 \\
        24.061 & 2.90 & 100.937 & 2.88 & 200.500 & 2.76 \\
        24.277 & 2.86 & 100.989 & 2.02 & 200.526 & 5.$67^\dagger$ \\
        24.339 & 5.29 & 101.017 & 1.98 & 200.546 & 2.27 \\
        24.461 & 2.37 & 101.027 & 2.81 & 200.560 & 2.94 \\
        24.680 & 2.34 & 101.082 & 2.14 & 200.573 & 2.91 \\
        24.764 & 3.$60^\dagger$ & 101.120 & 2.31 & 200.588 & 2.25 \\
        24.937 & 2.84 & 101.164 & 2.51 & 200.604 & 3.60 \\
        25.095 & 2.62 & 101.224 & 2.01 & 200.626 & 13.$18^*$ \\
        25.117 & 2.21 & 101.237 & 2.30 & 200.639 & 3.10 \\
        25.435 & 2.76 & 101.290 & 3.33 & 200.659 & 1.97
    \end{tabular}
    \caption{Inverse participation ratios of the Husimi plots of the
      first $36$ eigenstates in the low ($k=20$), medium ($k=100$) and
      high ($k=200$) energy ranges.  Bouncing-ball states are labeled with
      an asterisk, and states visibly scarred on the horizontal-bounce orbit
      are labeled with a dagger.  The random-matrix theory prediction,
      with bouncing-ball correction, is $2.44$, $2.29$ and $2.25$,
      respectively, in the three energy ranges.}
    \label{table:IPRs}
\end{table}
The average IPRs for the first 100 states in each energy regime are
summarized in Table~\ref{table:IPR-avgs}.
\begin{table}
    \begin{tabular}{ccccc}
        (a)              & (b)       & (c)
            & (d)            & (e) \\
        $k_{\text{min}}$ & IPR       & bb fraction of
            & IPR without bb & IPR without bb \\
                         & numerical & phase space
            & RMT prediction & numerical \\
        \hline
        20  & 3.67 (3.53) & 13\%  & 2.44 & 2.71 (0.62) \\
        100 & 3.60 (3.33) & 5.8\% & 2.29 & 2.70 (0.65) \\
        200 & 3.08 (1.98) & 4.1\% & 2.25 & 2.82 (0.80) 
    \end{tabular}
    \vspace{3.0ex}
    \caption{%
      Averages of the inverse participation ratios of the first 100
      states in energy ranges starting at $k = 20$, $100$ and $200$.
      Columns: (a) the $k$ value at the beginning of the range; (b)
      the actual IPR; (c) the estimated fraction of phase space
      occupied by the bouncing-ball states in this range of $k$; (d)
      the predicted IPR when the bouncing-ball region is excluded; (e)
      the computed average IPR after the bouncing-ball states (e.g.,
      those marked with an asterisk in Table~\ref{table:IPRs}) are
      omitted. Standard deviations are shown in parentheses. The large 
      remaining discrepancy between (d) and (e)
      is mostly due to symmetry effects, as explained in
      the text.}
    \label{table:IPR-avgs}
\end{table}
The IPRs cluster around 3, with some much larger.  The mean IPR is
3.67, 3.60 and 3.08 in the low, medium and high energy ranges
respectively, but with large standard deviations.  Visual examination
of the Husimi plots for states with the largest IPRs reveals that they
correspond to bouncing-ball states.  Since this phenomenon is well
understood (and is associated with marginally stable periodic orbits),
we remove this effect by removing the bouncing-ball states from the
average (see column (e) of Table~\ref{table:IPR-avgs}).

The remaining states avoid the bouncing-ball region, causing their IPR
to be increased by an amount that can be estimated.  At $k \approx
20$, the bouncing-ball region occupies about 13\% of phase space.
Assuming the non-bouncing-ball states to vanish identically in that
region, the random-matrix theory prediction for the average IPR
without the bouncing-ball states becomes $2.16 \times 1.13 =
2.44$.  The size of the
bouncing-ball region in phase space is energy-dependent (and must of course
go to zero in the $k \to \infty$ limit, in order
to satisfy the Shnirelman quantum
ergodicity condition). As found by
B\"acker {\it et al.}~\cite{Baecker} the number of bouncing-ball
states scales as $N_{bb}(E) = 0.2 E^{3/4}$ and thus the density of
bouncing-ball states, which is proportional to the area occupied by
them in phase space, scales as $E^{-1/4}$ or $k^{-1/2}$.  This scaling
law is in agreement with our numerical results.
The adjusted RMT predictions are shown in column (d) of
Table~\ref{table:IPR-avgs}.

As seen in Table~\ref{table:IPR-avgs}, the average IPRs with
bouncing-ball states removed are still well above the random-matrix
prediction in each energy range.  The IPRs were obtained by averaging
over 100 states; the statistical uncertainty in the means is thus
about $1/\sqrt{100}\approx 0.1$ of the standard deviation, or roughly
0.1, which is comparable to the differences in average IPR between
different energy ranges.  We see that the IPR does not increase
markedly with wavevector $k$, so the stadium is not an example of weak
quantum ergodicity as studied by Kaplan and Heller~\cite{KH} (this
makes sense, since the stadium is strongly chaotic with rapid mixing
throughout its phase space with the exception of the bouncing ball
region).  However, neither is there a marked trend towards the RMT
predictions as $k$ increases by a factor of ten.

Thus, we have uncovered systematic evidence of localization of
eigenfunctions in the stadium billiard beyond what would be predicted
on the basis of random-matrix theory.  What is responsible for it?
The next highest IPRs, after the bouncing-ball states, are for
eigenfunctions visibly scarred along the periodic orbit that bounces
horizontally between the centers of the endcaps (these
``horizontal-bounce states'' are marked with a dagger in
Table~\ref{table:IPRs}).
Thus, one might hypothesize that scarring along this and other
periodic orbits is partly responsible for the excess localization.
In the next Section, however, we shall
introduce a method that allows the excess localization to be
identified with specific classical structures in phase space, and
techniques to predict that localization theoretically.  We shall find
that quantum symmetry effects cause most of the excess localization, while
a secondary effect consists of combined scarring contributions from all of the
short unstable periodic orbits.

%%%%%%%%%%%%%%%%%%%%%%%%%%%%%%%%%%%%%%%%%%%%%%%%%%%%%%%%%%%%%%%%%%%%%%

\section{Spectral localization}
\label{locdens}

\subsection{Introduction}

The IPRs of individual eigenstates establish that there is
localization compared with the predictions of random matrix theory.
However, it does not permit the localization to be definitively
identified with structures in classical phase space because the IPR is
an aggregate over all of phase space.  But it is also possible to ask a
complementary question, namely, how ergodic are the eigenstates at a
given point in phase space as one varies the energy?  To answer this
question, we turn to the {\em local IPR}---a tool that provides a
picture of the regions of phase space that are prone to intensity
enhancement or
depletion over an ensemble of eigenstates.

\subsubsection{Definition}
\label{background}

The local IPR (`LIPR') is the mean squared eigenstate intensity at
point $(q,p)$, averaged over an ensemble of eigenstates~\cite{old}:
\begin{equation}
    \text{LIPR}(q,p) =
    \frac{\displaystyle\frac{1}{N} \sum_{n=1}^N
              \Bigl| \braket{q,p}{n} \Bigr|^4}
         {\displaystyle \biggl( \frac{1}{N}\sum_{n=1}^N
              \Bigl| \braket{q,p}{n} \Bigr|^2 \biggr)^2}
    \label{eq:LIPR-def}
      \,,
\end{equation}
where $N$ is the number of eigenstates being summed over.  Typically
the sum is taken over eigenstates in a small energy range around some
central value of $k$.  The Gaussian wavepackets $\ket{q,p}$ are
adjusted to maintain constant aspect ratio in phase space as $k$
changes (see discussion following Eq.~(\ref{eq:gaussian})); this keeps
the classical dynamics fixed as the energy increases.  Several LIPR
pictures are shown in Fig.~\ref{fig:LIPR}.

\subsubsection{Intuitive description}

The LIPR at the point $(q,p)$ is a statistical property of the set of
`random' variables $\{\braket{q,p}{n}: n = 1 \ldots N\}$.  The second
moment of this quantity [which appears in the denominator of
Eq.~(\ref{eq:LIPR-def})] is a smooth function of $q$ and $p$, as will
be shown in Sec.~\ref{sec:second-moment}; it is present as a
normalization factor.  The lowest moment that is nontrivial is the
fourth moment, which appears in the numerator of
Eq.~(\ref{eq:LIPR-def}).  The LIPR, then, measures the non-uniformity
of the local density of states at a (fixed) position in phase space.
A large LIPR indicates that a relatively small fraction of eigenstates
have large overlaps with the test Gaussian, whereas a small LIPR
indicates that the overlaps are distributed in a more `egalitarian'
way.

We expect the LIPR to be a sensitive indicator of the presence of
scarring, because a wavepacket centered on a periodic orbit has a
local density of states that oscillates with energy~\cite{H1}.
Eigenstates in the peak of the oscillation will, on average, have
enhanced overlaps with the test state, while eigenstates in the trough
will have suppressed overlaps.  This nonuniformity will cause the LIPR
to have peaks near periodic orbits.

There is a useful interpretation of the LIPR---it is proportional to
the long time average of the probability that the wavepacket $\ket{q,p}$,
evolved in time, has returned to its original location.  A derivation
and discussion of this interpretation are presented in
Sec.~\ref{sec:return-prob} below.

Since the stadium is strongly chaotic, classical trajectories explore
all of phase space (except for a set of measure zero).  Therefore the
classical return probability for long times is uniform, and the
resulting naive classical prediction is that $\text{LIPR} = 1$.  This
is indeed too naive.

Instead, the `null hypothesis' for a chaotic system is that its eigenfunctions
are random superpositions of plane waves~\cite{berry}.  Under that
maximally random assumption, projections of wavefunctions on test
states should follow a $\chi^2$ distribution with one degree of
freedom (if the eigenfunctions and test functions are real) or a
$\chi^2$ distribution with two degrees of freedom (if they are
complex).  The corresponding predicted LIPRs are 3 or 2, respectively---already
different from the naive classical prediction.

\subsection{Introduction to pictures}

In Fig.~\ref{fig:LIPR} we present the LIPR computed for even-even
eigenstates in the wavenumber ranges $k=50$ to $60$, $100$ to $150$
and $200$ to $225$.  They display interesting and beautiful
localization effects.

The horizontal axis of each picture is $q$, the position
along the boundary, which runs from $0$ to
$L$.  $q=0$ corresponds to the midpoint of one of the straight walls.
The vertical axis is the component of momentum parallel to the boundary
$p/k$, which runs from $-1$ to $1$.  The
eightfold symmetry of these plots results from the two spatial
reflection operations plus time-reversal symmetry.  The LIPRs all show
a very large amplitude in the bouncing-ball region, explainable by the
bouncing-ball and near-bouncing-ball states present in the energy
ranges considered.  But in addition, there is localization in many
other regions of phase space.  There are very prominent straight lines
of enhanced LIPR running along the lines of symmetry, and filamentary
streaks running across other parts of phase space.  There are also
isolated spots at which the LIPR is unusually large.

Note that, although the details vary somewhat, the main localization
features appear in the same positions in each energy range.  This lack
of $\hbar$-dependence suggests that the streaks are associated with
classical phenomena.  Our main goal is to give a quantitative
explanation for all of these effects.  We would also like to
understand the relative contributions of each semiclassical effect to the
total average IPR enhancement found in Sec.~\ref{sec:locprops} above.

If one varies the aspect ratio of the test state to make it more
position-like or momentum-like, a series of bright spots can also be
resolved at various points along the streaks, on a scale as small as
the smaller axis of the Gaussian test state (see
Fig.~\ref{fig:LIPR-aspect}).

\subsection{Mean wavefunction intensity}
\label{sec:second-moment}

The mean wavefunction intensity appears in the denominator of
Eq.~(\ref{eq:LIPR-def}) in the role of a normalization factor.  It can
be determined from classical phase space arguments similar to those
used in the derivation of Weyl's law.  It depends on $p$ for two
reasons: (1) we are using normal derivatives, which introduces a
contribution of $p_\perp = k \sqrt{1 - (p/k)^2}$ multiplying the
coefficients of the plane waves (since the coefficients are squared,
the resulting factor is $k^2(1 - (p/k)^2)$), and (2) there is a geometrical
factor coming from the projection from the circle $k_\perp^2 + p^2 =
k^2$ onto the boundary of the stadium.  Because the plane waves are
evenly distributed around the circle, there are more near $p = k$ than
near $p = 0$; the resulting geometrical factor is $\left[ 1 - (p/k)^2
\right]^{-1/2}$.  Together, (1) and (2) give
\begin{equation}
    \avg{ \Bigl| \braket{q,p}{n} \bigr|^2}_n \sim
    \sqrt{1-(p/k)^2}
    \label{eq:mean-intensity}
    .
\end{equation}
This smooth factor is divided out of the LIPR in
Eq.~(\ref{eq:LIPR-def}), giving the LIPR a flat
background.

We checked the mean wavefunction intensity for the computed even-even
states in the interval from $k=100$ to $110$.  It agrees well with
Eq.~(\ref{eq:mean-intensity}), except at $p = 0$ around the centers of
the straight segments and the centers of the endcaps, where sharp
peaks are seen.  The peaks occur because only even-even states have
been included in the analysis; since the odd states must vanish at the
symmetry points, the even-even states compensate by having double the
average intensity there.  Apart from this, the agreement of the
computed mean intensity with Eq.~(\ref{eq:mean-intensity}) indicates
that we have included enough eigenfunctions in the energy ranges
chosen for Fig.~\ref{fig:LIPR}.

\subsection{Symmetry lines}
\label{sec:rolesym}

What is the explanation for the streaks in the LIPR plots of
Fig.~\ref{fig:LIPR}?  The most prominent streaks are shown
schematically in Fig.~\ref{fig:symmetry-structures}.  The streak
labeled (1) in Fig.~\ref{fig:symmetry-structures} corresponds to
trajectories emerging from the center of the endcap.  The streak
labeled (2) corresponds to a family of orbits emerging from the center
of the straight segment.  Streak (3), which starts at small momentum
at the edge of the bouncing-ball region and extends to large momentum
at the center of the endcap, corresponds to orbits that have a
vertical segment in the endcap region.  The box and the bowtie are the
two most prominent periodic orbits in this family.  The streak labeled
(4) starts at small momentum at the center of the straight segment,
curves up to larger momentum, and finally comes back down to zero
momentum at the center of the endcap.  It corresponds to orbits that
pass through the center of the stadium.  The most prominent periodic
orbit in this family is the Z orbit.  Finally, there is the horizontal
line (5) going across the plot at zero momentum, corresponding to
normal incidence on the billiard boundary.  All of these families can
be understood by considering time reversal and parity symmetry effects in
combination with dynamics, as we will now see.

\subsubsection{Simple explanation}
\label{sec:simple}

The intuitive reason for the symmetry lines is that the system, having
time-reversal symmetry, has real eigenfunctions.  However, the
Gaussian test states [Eq.~(\ref{eq:gaussian})] are intrinsically
complex.  Therefore the overlap $\braket{q,p}{n}$ has both real and
imaginary parts which, for most choices of $q$ and $p$, are not
trivially related to one another.  Therefore the typical computed LIPR
for the system is 2, characteristic of complex eigenfunctions.  But in
a sense the `correct' value $\text{LIPR} = 3$ (characteristic of real
eigenfunctions) is obscured by a non-ideal choice of test states
whose properties do not match those of the eigenstates.

However, due to the symmetry of the system, there are certain values
of $q$ and $p$ for which the real and imaginary parts of
$\braket{q,p}{n}$ are not random and independent. For example, when $p = 0$
(normal incidence to the wall), the test states and their projections
$\braket{q,p=0}{n}$ on the eigenstates become pure real and therefore
near $p = 0$ the LIPR increases to 3.  [This explains the symmetry
line labeled (5) in Fig.~\ref{fig:symmetry-structures}.]  Similarly, the
eigenfunctions are symmetric about the centers of the straight walls,
and therefore $\braket{q=0,p}{n}$ is pure real and again the LIPR
increases to 3.  (Similarly for the centers of the endcaps.)  This
simple argument explains the straight symmetry lines (1), (2) and (5)
that appear in
the LIPR pictures.  A rigorous derivation of their heights and widths
will be given in the next Section.

It is interesting to note that the same LIPR enhancements would be
present if we had used eigenstates of other symmetry classes, or put
eigenstates of all symmetry classes together.  For
example, for odd-odd eigenstates, the eigenfunctions are antisymmetric
about the centers of the straight walls, and therefore
$\braket{q=0,p}{n}$ is pure imaginary, again giving a $\chi^2$
distribution with one degree of freedom, and the enhancement
$\text{LIPR} = 3$.

\subsubsection{Derivation}
\label{structsym}

In this Section we show how to compute the enhancement of the LIPR that
appears near the symmetry lines of the system.  We do this in order to
verify the conclusions of the preceding simple arguments, and also 
to deduce important
details such as the profiles and widths of the symmetry lines.

The derivation relies on the fact that the eigenfunctions are chosen to
be be real (in a coordinate basis), which can be done as a consequence
of time-reversal symmetry.  We denote time-reversal with $T$, an
antiunitary operator.  On the surface of section, $T\ket{\p++}
= \ket{\p+-}$.

Furthermore, the eigenfunctions are symmetric or antisymmetric with
respect to the reflections $x \rightarrow -x$ and $y \rightarrow -y$,
which we will denote with the unitary operators $R_x$ and $R_y$
respectively.  On the surface of section, this corresponds to symmetry
with respect to reflection about two values $\{q_x,q_y\}$
corresponding to an intersection of the billiard wall with the $y$ and
$x$ axes, respectively.  Then for example, $R_x \ket{q_x+\Delta q,p} =
\ket{q_x-\Delta q,-p}$ (where $q$ is always taken modulo the billiard
perimeter).
For simplicity in the following derivation we will consider the
simpler case of a single reflection symmetry, $R$, about $q=0$.

It is not correct to model the eigenfunctions as uncorrelated Gaussian
random variables, because their reflection symmetry correlates their
values at different points.  However, we can generate random
wavefunctions $\ket{n,\pm}$ with positive (`$+$') or negative (`$-$')
symmetry by taking completely random real wavefunctions $\ket{n}$
(with no symmetry) and projecting them onto the correct symmetry
subspace:
\begin{equation}
    \ket{n,\pm} = \recip{\sqrt{2}} \left( \ket{n} \pm R \ket{n} \right)
    \label{eq:symmetrize}
    .
\end{equation}

To substitute into Eq.~(\ref{eq:LIPR-def}),
we will need to compute $\braket{\p++}{n,\pm}$.  This quantity is
complex, so we decompose it into two real random variables $\mu_{\pm}$
and $\nu_{\pm}$ as
\begin{equation}
    \recip{\sqrt{2}} \left[
        \braket{\phi_{q,p}}{n} \pm \matrixel{\phi_{q,p}}{R}{n}
    \right] \equiv \mu_{\pm} + i \nu_{\pm}
    .
\end{equation}
If $\mu_{\pm}$ and $\nu_{\pm}$ had the same variances and were
uncorrelated, the LIPR would uniformly equal $2$, like that of any
Gaussian random process with two degrees of freedom.  This null result
would hold even if the variance depended on $p$ and $q$, 
because the square of the
variance appears both in the numerator and in the denominator of the
definition of the LIPR.
 
However, in reality the symmetries cause the variances of the real and
imaginary parts $\mu_{\pm}$ and $\nu_{\pm}$
to depend differently on phase space location.
The effective number
of degrees of freedom varies from two (when $\langle\mu_{\pm}^2\rangle
= \langle\nu_{\pm}^2\rangle$) to one (when $\langle\mu_{\pm}^2\rangle
\neq 0$, $\langle\nu_{\pm}^2\rangle = 0$), and correspondingly the LIPR
varies from $2$ to $3$.

In terms of the three quantities $\langle\mu_{\pm}^2\rangle$,
$\langle\nu_{\pm}^2\rangle$, and $\langle\mu_{\pm}\nu_{\pm}\rangle$,
\begin{equation}
    \avg{\abs{\mu_{\pm} + i \nu_{\pm}}^2} =
    \savg{\mu_{\pm}^2} + \savg{\nu_{\pm}^2}
    ,
\end{equation}
and $\avg{\abs{\mu_{\pm} + i \nu_{\pm}}^4}$ can be expanded and then
contracted pairwise, giving
\begin{equation}
    \avg{\abs{\mu_{\pm} + i \nu_{\pm}}^4} =
    2 \left( \savg{\mu_{\pm}^2} + \savg{\nu_{\pm}^2} \right)^2 +
    \left( \savg{\mu_{\pm}^2} - \savg{\nu_{\pm}^2} \right)^2
      + 4 \savg{\mu_{\pm}\nu_{\pm}}^2
\label{munumoments}
\end{equation}
The problem is reduced to the computation of the required variances
and correlations in Eq.~(\ref{munumoments}).

Using $\braket{\p++}{n}^* = \braket{\p+-}{n}$ and $R \ket{\p++} =
\ket{\p--}$, we obtain
\begin{eqnarray}
    \mu_{\pm} & = &
    \recip{2\sqrt{2}}\left[
        \braket{\p++}{n} + \braket{\p+-}{n}
        \pm \braket{\p--}{n} \pm \braket{\p-+}{n}
    \right] \\
    \nu_{\pm} & = &
    \recip{2i\sqrt{2}}\left[
        \braket{\p++}{n} - \braket{\p+-}{n}
        \pm \braket{\p--}{n} \mp \braket{\p-+}{n}
    \right]
    .
\end{eqnarray}
Because $\ket{n}$ represents not a wavefunction but rather the
projection of a wavefunction onto the surface of section, the closure
relationship does not hold: $\sum_n \ketbra{n}{n} \ne \openone$.  But
as explained above in Sec.~\ref{background}, it is still approximately true
that
\begin{equation}
    \sum_n | n \rangle \langle n | \p++ \rangle \approx
    f(p) \ket{\p++}
    ,
\end{equation}
where $f(p) = f(-p) \sim \sqrt{1 - (p/k)^2}$.  It follows that
\begin{mathletters}
    \begin{eqnarray}
        \avg{\mu_{\pm}^2} & = &
        \recip{8N} \sum_n
        \Bigl[ \braket{\p++}{n} + \braket{\p+-}{n}
        \pm \braket{\p--}{n} \pm \braket{\p-+}{n} \Bigr]
        \nonumber \\
        & & \hphantom{\recip{8N} \sum_n}
        \times \Bigl[ \braket{n}{\p++} + \braket{n}{\p+-}
        \pm \braket{n}{\p--} \pm \braket{n}{\p-+} \Bigr]
        \nonumber \\
        & = & \frac{f(p)}{2N}
        \Bigl[ \pp++ + \Real\pp+- \pm \Real\pp-+ \pm \pp-- \Bigr]
        ;
    \end{eqnarray}
    the last line follows from $\braket{\p+-}{\p-+} = \pp--$, which is
    real.  The other correlations can be worked out similarly; the
    results are
    \begin{eqnarray}
        \avg{\nu_{\pm}^2} & = &
        \frac{f(p)}{2N}
        \Bigl[ \pp++ - \Real\pp+- \mp \Real\pp-+ \pm \pp-- \Bigr] \\
        \avg{\mu_{\pm}\nu_{\pm}} & = &
        \frac{f(p)}{2N} \Bigl[ \Imag\pp+- \pm \Imag\pp-+ \Bigr]
        .
    \end{eqnarray}
\end{mathletters}

\subsubsection{Even and odd subsets vs. whole set}

At this point we must distinguish between two types of LIPR.  First,
one can evaluate the LIPR by averaging (in the numerator and the
denominator) over only the even states (as done in this paper) or only
the odd states.  In that case one obtains
\begin{mathletters}
    \label{eq:LIPR1}
    \begin{eqnarray}
        \text{LIPR}(q,p; \{\ket{n,\pm}\})
        & = & 2 +
        \frac{\abs{\pp+- \pm \pp-+}^2}{\left(\pp++ \pm \pp-- \right)^2} \\
        & = & 2 +
        \frac{\abs{\pqavg{T} \pm \pqavg{RT}}^2}
        {\left(\braket{\p++}{\p++} \pm \pqavg{R} \right)^2}
        .
    \end{eqnarray}
\end{mathletters}
On the other hand, one can evaluate the LIPR by averaging over both the
even and the odd states.  In that case,
\begin{mathletters}
    \label{eq:LIPR2}
    \begin{eqnarray}
        \text{LIPR}\left( q,p; \{\ket{n,+}\} \cup \{\ket{n,-}\} \right)
        & = & 2 +
        \frac{\abs{\pp+-}^2 + \abs{\pp-+}^2}{\pp++^2 + \pp--^2} \\
        & = & 2 +
        \frac{\abs{\pqavg{T}}^2 + \abs{\pqavg{RT}}^2}
        {\braket{\p++}{\p++}^2 + \pqavg{R}^2}
        .
    \end{eqnarray}
\end{mathletters}
The required matrix elements can be worked out from
Eq.~(\ref{eq:gaussian}):
\begin{equation}
    \abs{\braket{q,p}{q',p'}} =
    \exp\left[
        - \frac{(q - q')^2}{4 \sigma^2} - \frac{\sigma^2 (p - p')^2}{4}
    \right]
    .
\end{equation}

The three LIPRs---Eq.~(\ref{eq:LIPR1}) for even or odd symmetry and
Eq.~(\ref{eq:LIPR2}) for both symmetry classes put together---all
can be shown to give results between $2$
and $3$.  In fact, outside the region near $q = p = 0$ the three
versions are identical; they all predict bright symmetry lines near $q
= 0$ (or near any $q$ associated with a parity symmetry) of the form
\begin{equation}
    \text{LIPR}(q, \abs{p} \gg 1/\sigma) =
    2 + e^{-2 q^2/\sigma^2}
    \label{eq:sym-q=0}
\end{equation}
and bright symmetry lines near $p = 0$ of the form
\begin{equation}
    \text{LIPR}(\abs{q} \gg \sigma, p) =
    2 + e^{-2 p^2 \sigma^2}
    \label{eq:sym-p=0}
    .
\end{equation}
We note that the point $p=q=0$, near which the three LIPR definitions given
above do not agree, will always be on a short periodic orbit (e.g.,
the horizontal bounce orbit of the stadium billiard), and so the behavior
there in any case cannot be determined without taking dynamical scar effects
into account (see Section~\ref{scarsection}).

\subsubsection{Curved symmetry lines}
\label{sec:curved}

The curved streaks (3) and (4)
also correspond to classical structures related to
the system's symmetry.  Consider first the streak labeled (3) in
Fig.~\ref{fig:symmetry-structures}.  It corresponds to trajectories with a
vertical segment in the endcap region.  In the desymmetrized
quarter-stadium, these trajectories are incident
normally on the $y=0$ horizontal boundary, so a Gaussian test state
with zero tangential momentum placed at the point of normal incidence
$(q',p=0)$
would give rise to an
enhanced LIPR of 3, as for the zero-momentum line (5) above. Of course,
this point of normal incidence exists only on the boundary
of the {\it quarter}-billiard. At other
points along the same orbit, including points that also live on the boundary
of the full stadium, a Gaussian with momentum aligned along
the trajectory will be close to, but not exactly equal to,
a time iterate of this zero-momentum
Gaussian living on the $y=0$ boundary. The exact time iterate of
this purely real Gaussian will in general be a Gaussian centered on
the other periodic point $(q,p)$ but with a (possibly complex) width $\sigma$
somewhat different from that used in our test state.  Thus, our test
Gaussian at $(q,p)$ 
will have significant overlap with an iterate of a purely
real state at $(q',p=0)$
and result in an enhanced LIPR somewhere between 2 and 3. Very
little wavepacket rotation or stretching occurs during the short vertical
trip between the $y=0$ line and the endcap; it is for this reason that the
streak (3) is so strong. Further iterations of the dynamics will result in
additional streaks of enhanced IPR; these however will be much weaker due to 
the additional stretching which makes a circular wavepacket centered at these
distant points less closely related to the evolved version of the
$(q',p=0)$ real
wavepacket. All streaks in the LIPR plots which have not been 
explicitly identified in Fig.~\ref{fig:symmetry-structures} may be explained
as further iterates of symmetry lines we have discussed explicitly.
The rather strong streak (4), associated with trajectories going through
the center of the stadium, has a completely analogous explanation.

\subsubsection{Quantitative contribution of the symmetry lines to the IPR}

We can now estimate the contribution of the symmetry lines to the
predicted average IPR.  There are twelve strong
roughly vertical streaks (including
the curved streaks (3) and (4)) in
the region $p \ne 0$ and one horizontal streak at $p=0$.  We assume
temporarily the curved symmetry lines to have a central height
near 3 and width of order $\sigma$ just as for the $x=0$ symmetry
line.  Integrating the area under these streaks assuming
Eqs.~(\ref{eq:sym-q=0}) and (\ref{eq:sym-p=0}) gives a predicted IPR
from symmetry effects alone of 2.51 (in the energy range from $k =
100$ to $150$).  
The enhancement of the LIPR in the whispering-gallery
region and in bright spots due to periodic orbits gives a contribution
only of about 0.03, negligible in comparison to the symmetry effects.
Since the bouncing-ball region, which covers about 6\% of phase space,
is excluded in the above analysis, we should increase the prediction
by 6\% to 2.66 for the purpose of comparison with the numerical value
of $2.70 \pm 0.07$.
In the energy range from $k=200$ to $225$ the corresponding
predicted IPR is 2.40, or 2.51 corrected for the bouncing-ball region,
whereas the numerical value is $2.82 \pm 0.08$ (the discrepancy in the
higher-energy region remains unexplained).
Thus, it seems that inclusion of symmetry and bouncing ball effects is
sufficient to yield a quantitative understanding of much of
the excess average
IPR in Table~\ref{table:IPR-avgs}.
Interestingly, scarring plays little role
in the excess average IPR, but a very important role in understanding
the phase-space structure.

\subsubsection{LIPR using real wavepackets}
\label{sec:LIPR-desym}

In light of our claim in Sec.~\ref{sec:simple} that complex Gaussian test
states living on the boundary of the billiard
are not ideal for probing eigenstates in a system with time
reversal symmetry, it is
natural to ask what test states would be more appropriate. A natural
approach would be to symmetrize the test wavepackets with respect to
time reversal symmetry, by taking their real or imaginary parts. Equivalently,
since the eigenstates are chosen to be pure real, one may simply take
the real or imaginary parts of the overlaps 
$\braket{q,p}{n}$. We note that one might also consider symmetrizing the
test Gaussians with respect to parity symmetry, which is also present in
our system. This, however, would have no effect on any observed quantities,
since all of the eigenstates already respect parity symmetry.
We then define 
\begin{equation}
    \text{LIPR}_{\text{real}}(q,p) = 
    \frac{\displaystyle\frac{1}{N} \sum_{n=1}^N
              \Bigl| \text{Re} \braket{q,p}{n} \Bigr|^4}
         {\displaystyle \biggl( \frac{1}{N}\sum_{n=1}^N
              \Bigl| \text{Re} \braket{q,p}{n} \Bigr|^2 \biggr)^2} \,.
    \label{eq:LIPR-desym-def}
\end{equation}
This quantity is plotted in
Fig.~\ref{fig:LIPR-desym}, and should be compared with the corresponding
pictures in Fig.~\ref{fig:LIPR} for the LIPR of the original complex
wavepackets. Because the 
new test states are real,
the background LIPR rises from 2 to 3,
while the LIPR along the straight symmetry lines remains unchanged (at 3).

Naively, one might expect the desymmetrization process described here to
completely eliminate all symmetry effects, leading to flat
$\text{LIPR}_{\text{real}}(q,p)=3$ behavior, with the exception of
dynamics-related enhancement in the bouncing-ball region and  near isolated
unstable periodic orbits. Instead, one finds that although the straight
symmetry lines ($p=0$ and the parity symmetry lines) have indeed disappeared
in Fig.~\ref{fig:LIPR-desym}, $\text{LIPR}_{\text{real}}$ along the
curved symmetry lines (e.g. the streaks identified as (3) and (4) in
Fig.~\ref{fig:symmetry-structures})
is enhanced from $3$ to $\approx 4.5$, making
these curved streaks just as visible as in the original plot of
Fig.~\ref{fig:LIPR}. This is easy to understand based on our analysis
of these streaks in Sec.~\ref{sec:curved}. We saw there that wavepackets
located on the curved streaks are time-evolved versions of real wavepackets.
So let $r_n$ be the overlap of eigenstate $|n\rangle$
with a real wavepacket (i.e.
either one having $p=0$ or one that lives on a parity symmetry line). $r_n$
of course is distributed as a real Gaussian variable. Now the overlap of
the same eigenstate $|n\rangle$ with a time-evolved copy of the real wavepacket
will be given by $r_n'=r_n \exp{i \theta_n}$, where in the semiclassical
regime $\theta_n$ can be taken to be a random phase. The LIPR for $r_n'$
is of course equal to that of $r_n$ (i.e., $3$),
 as the phase has no effect on the
intensity $|r_n'|^2=|r_n|^2$. If, however, we take the real part
of the wavepacket
living on the curved symmetry line,
we must look at the quantity
\begin{equation}
\text{LIPR}_{\text{real}}= {{1 \over N} \sum_{n=1}^N (\text{Re} \;r_n')^4
\over \left(  {1 \over N} \sum_{n=1}^N (\text{Re} \;r_n')^2 \right)} =
{9 \over 2}\,
\end{equation}
because $\text{Re} \; r_n'=r_n \cos {\theta_n}$,
where $r_n$ is Gaussian-distributed
and $\theta_n$ is a random phase. This is in contrast with the LIPR
of $3$ obtained on the straight symmetry lines as well as in the background
(the LIPR on the straight symmetry lines would also be enhanced to $9/2$
if we had let the eigenstates have random phases instead of adopting the
convention where they are all real). Thus, there appears to be no natural
and simple way to eliminate all symmetry effects on the LIPR (and thus,
also on the IPR) in this
time-reversal
invariant system.

\subsection{Dynamics}

In this Section we will consider the LIPR for a system with no
time-reversal symmetry, where the wavefunctions are complex instead of
pure real, and the arguments of Section~\ref{sec:rolesym} concerning
symmetry lines do not apply. In such a system the LIPR plot would be
far more uniform, and most of the remaining structure would consist of
isolated regions of enhanced LIPR, against a background of
$\text{LIPR}=2$ (corresponding to the statistics of the square of a
complex Gaussian variable).  The streaks would be all gone, including
the vertical parity-line streaks, which also depend on time-reversal
symmetry for their existence.  Thus in this Section we study the
effect of dynamics alone on the LIPR.  First we give a general theory
for computing the LIPR classically anywhere in phase space, and then
we show an easier way to compute the LIPR in the neighborhood of a
periodic orbit.  In the process, we will also see how to treat the
case of a periodic orbit that happens to lie on a symmetry line in a
time-reversal invariant system.  In the following Section
(Sec.~\ref{sec:dynamsym}), we will describe an alternative way to
treat parity symmetry and dynamical effects in a unified framework, to
better understand scar effects in a system with symmetry.

\subsubsection{Full dynamics treatment}

\label{dynamipr}

In this Section we show how to compute the enhancement of the LIPR due
to the short-time dynamics of the system.  Since we are considering the
surface-of-section wavefunctions, the classical dynamics that is
important is the classical Poincar\'e map $\MCl$.  However,
since the effects we are seeking are quantum mechanical, we need to
start from the corresponding {\em quantum\/} Poincar\'e map.  For this
we use the operator $M(k)$ that enters into the boundary integral
method:
\begin{equation}
    \matrixel{q}{M(k)}{\phi} \equiv
      \oint dq' K_{\text{ex}}(q, q'; k) \braket{q'}{\phi}\,,
    \label{eq:BIM-map}
\end{equation}
where the kernel is \cite{Boasman}
\begin{equation}
    K_{\text{ex}}(q, q'; k) \equiv
    -\frac{ik}{2} H_1^{(1)}(k \abs{\bvec{r}(q) - \bvec{r}(q')})
    \; \vhat{n}(q) \cdot [\bvec{r}(q) - \bvec{r}(q')]
    \label{eq:Kex}
    ,
\end{equation}
$\bvec{r}(q)$ is a point on the billiard wall at arc length $q$, and
$\vhat{n}(q)$ is the inward-pointing unit normal vector at $q$. Does
$M(k)$ as defined generate the correct quantum dynamics?  In a sense
that is a philosophical question since the Poincar\'e map has no
obvious exact quantum counterpart.  However $M(k)$ satisfies enough
properties expected of a quantum Poincar\'e map that we shall use it:
\begin{enumerate}
\item To within a semiclassical approximation it agrees with the
    semiclassical Poincar\'e map \cite{Boasman} (which is easier to
    define~\cite{Bogomolny}).
\item It is approximately unitary within a subspace whose dimension is
    given by the area of the Poincar\'e section in units of Planck's
    constant, and exponentially small beyond that dimension.
\item If a boundary function $\ket{n}$ corresponds to a wavefunction
    that satisfies the boundary conditions at wavenumber $k = k_n$,
    then $\ket{n}$ is mapped onto itself under the action of $M(k)$:
    \begin{equation}
        M(k_n)\ket{n} = \ket{n}
        .
    \end{equation}
    Indeed, this is the boundary integral method criterion for an
    eigenstate.
\end{enumerate}

The following derivation should be considered suggestive rather than
rigorous.  For example, $M(k)$ is approximately unitary but it has
actual eigenvalues both inside and outside the unit circle.  This
leads to difficulties analytically continuing it to the region of
interest.

Given properties 2 and 3 above, it is clear that one crude way to find
surface eigenstates of the system would be to apply $M(k)$ repeatedly
to an arbitrary initial state $\ket{\phi}$ and average the results:
\begin{eqnarray}
    \ket{n} & \propto & \lim_{\epsilon \rightarrow 0^+}
    \sum_{j=0}^{\infty} [M(k_n) - \epsilon]^j \ket{\phi} \\
    & = & G(k_n) \ket{\phi},
\end{eqnarray}
where
\begin{equation}
    G(k) \equiv \lim_{\epsilon \rightarrow 0^+}
    \recip{1 - M(k) + \epsilon}
\end{equation}
is similar to a Green's function.  Since the component of $\ket{\phi}$
that projects onto $\ket{n}$ is unchanged by the action of $M(k_n)$,
whereas the other components are multiplied by a complex phase, the
only component that is not averaged out by the above procedure is the
projection on $\ket{n}$.

Whenever $k$ passes through an eigenvalue $k_n$, one of the
eigenvalues $e^{i \alpha_n(k)}$ of $M(k)$ passes along the unit circle
through $1$ and $G(k)$ has a singularity.  The singular part of $G$
can be written schematically as
\begin{equation}
    \recip{\pi} \Real G(k) = \sum_n
    \frac{\ket{n} \delta(k - k_n)\bra{n}}
    {\abs{d\alpha_n(k)/dk}}
    .
\end{equation}

The denominator gives the velocity of rotation of the $n\/$th
eigenvalue as it passes through $1$.  It can be shown from
semiclassical arguments that the average of these velocities is given
by the average length of the trajectory segments corresponding to one
Poincar\'e map:
\begin{equation}
    \avg{\frac{d\alpha_n}{dk}}_n = \avg{\ell} = \pi \frac{A}{L}
    \label{eq:dalpha-dk}
    .
\end{equation}
The second equality follows from a classical theorem well known in
acoustics~\cite{avg-len}, where $A$ is the billiard's area and $L$ the
length of its perimeter.  For a chaotic system,
Eq.~(\ref{eq:dalpha-dk}) holds not only in an average sense but also
holds approximately for individual eigenphases.  Therefore we take
$\avg{\ell}$ outside of the summation, yielding the useful expression
\begin{equation}
    \frac{A}{L} \Real G(k) \approx
    \sum_n \ket{n} \delta(k - k_n)\bra{n}
    .
\end{equation}
Note that the sum is over states of the full system, each with its
distinct eigenvalue $k_n$.

It is a simple warm-up exercise to compute the denominator of the LIPR:
\begin{eqnarray}
    \sum_n \abs{\braket{q,p}{n}}^2
    & = & \sum_n \braket{q,p}{n}\braket{n}{q,p} \\
    & = & \sum_n \braket{q,p}{n} \int dk \delta(k - k_n) \braket{n}{q,p} \\
    & = & \frac{A}{L} \int dk \Real \matrixel{q,p}{G(k)}{q,p} \\
    & = & \frac{A}{L} \Real \int dk \left[
        \braket{q,p}{q,p} +
        \matrixel{q,p}{M(k)}{q,p} +
        \matrixel{q,p}{M^2(k)}{q,p} +
        \cdots
    \right]
    .
\end{eqnarray}
Now note that $M(k)$ involves phases that vary rapidly with $k$;
therefore only the first term in the brackets survives the integral
over $k$.  We are left with
\begin{equation}
    \sum_n \abs{\braket{q,p}{n}}^2
    \approx \frac{A \Delta k}{L} \braket{q,p}{q,p}
    ,
\label{secondmom}
\end{equation}
where $\Delta k$ is the range of $k$ over which the averaging is done.

To compute the fourth moment, we proceed as follows:
\begin{eqnarray}
    \sum_n \abs{\braket{q,p}{n}}^4 &=& 
    \sum_n \braket{q,p}{n} \braket{n}{q,p} \braket{q,p}{n} \braket{n}{q,p} \\
    &=& \sum_{n,m}
        \braket{q,p}{n} \int \! dk  \, \delta(k  - k_n) \braket{n}{q,p}
        \braket{q,p}{m} \int \! dk' \, \delta(k' - k_m) \braket{m}{q,p}
        \delta_{nm}.
\end{eqnarray}
Because of the {$\delta(k-k_n)$} and {$\delta(k'-k_m)$} factors, the
Kronecker delta {$\delta_{nm}$} may be replaced by {$C\delta(k-k')$}
where {$C$} is an undetermined constant with dimensions of momentum
which is needed to make its product with the delta function
dimensionless. Continuing, then, we have
\begin{eqnarray}
    \sum_n \abs{\braket{q,p}{n}}^4
    &=& C \left( {A \over L} \right)^2 \int dk \, dk'
        \Real \matrixel{q,p}{G(k)}{q,p}
        \Real \matrixel{q,p}{G(k')}{q,p} \delta(k-k')
    \\
    &=& C \left( {A \over L} \right)^2 \int dk \Bigl[
        \Real \matrixel{q,p}{G(k)}{q,p} \Bigr]^2 \\
    &=& C \left( {A \over L} \right)^2 \int dk \sum_{j,j'}
        \Real \langle q,p | M^j(k) | q,p \rangle 
        \Real \langle q,p | M^{j'}(k) | q,p \rangle \\
    &\approx& C \left( {A \over L} \right)^2 \int dk \sum_j \left[
        \Real \langle q,p | M^j(k) | q,p \rangle \right]^2
    \label{eq:fourthmom-1} \\
    &\approx& {C \over 2}
    \left( {A \over L} \right)^2 \int dk \sum_j \left|
        \langle q,p | M^j(k) | q,p \rangle \right|^2
    \label{eq:fourthmom-2} \,.
\end{eqnarray}
where in going to Eq.~(\ref{eq:fourthmom-1}) we have used the fact
that for $j \ne j'$, the two phases in the integrand are approximately
uncorrelated, and the integral averages to zero.
Eq.~(\ref{eq:fourthmom-2}) (in obtaining which we have used the fact
that for $j \ne 0$ the phase of $\langle q,p | M^j(k) | q,p \rangle$
is a rapidly changing function of $k$, and so may be regarded as
random in the integration; the $j=0$ term is irrelevant in the
infinite sum) may be interpreted as expressing the proportionality
between the LIPR and the mean quantum return probability, averaged
over {\it all} times. As a result of the reloading effect {\cite{KH2}}
the mean long-time return probability is proportional to the sum of
the {\it short-time} recurrences, which may be cut off at the mixing
time, $j_{\text{mix}} \equiv T_{\text{mix}}/T_B$:
\begin{equation}
\sum_n \abs{\braket{q,p}{n}}^4
    \approx {C' \over 2}
     \left( {A \over L} \right)^2 \int dk
      \sum_{j=-j_{\rm mix}}^{j_{\rm mix}} \left|
        \langle q,p | M^j(k) | q,p \rangle \right|^2 \,.
\label{fourthmomq}
\end{equation}
The constant of proportionality changes in going from
Eq.~(\ref{eq:fourthmom-2}) to Eq.~(\ref{fourthmomq}), as is reflected
in our change of notation from $C$ to $C'$. The mean quantum return
probability is inversely proportional to the number of Planck-sized
cells in phase space, or inversely proportional to the Heisenberg
time, and thus $C'$ is inversely proportional to the Heisenberg time.

But at times shorter than the mixing time, the quantum map $M^j(k)$
can be approximated semiclassically in terms of the classical
Poincar\'e map $\MCl^j$ (which does not depend on $k$) times a phase
that varies rapidly with $k$:
\begin{equation}
    \matrixel{q,p}{M^j(k)}{q,p} \approx
    \matrixelCl{q,p}{\MCl^j}{q,p}^{1/2} \exp(i k \ell_j)
    ,
\end{equation}
where $\ketCl{q,p}$ represents a {\em classical\/} Gaussian
probability distribution centered at $(q,p)$ (notationally
distinguished from a quantum wavefunction by the round brackets),
$\MCl^j$ is the classical Poincar\'e map iterated $j$ times, and
$\ell_j$ is the length of the corresponding classical trajectory.
$\matrixelCl{q,p}{\MCl^j}{q,p}$ is the classical overlap of the
original distribution with the $j$-times iterated distribution:
\begin{equation}
    \matrixelCl{q,p}{\MCl^j}{q,p} \equiv
        \int dq' \, dp' P_{\MCl^j(q,p)}(q',p') P_{q,p}(q',p')
    \label{eq:Cl-overlap} \,.
\end{equation}
So finally we have
\begin{equation}
\sum_n \abs{\braket{q,p}{n}}^4 \approx
    {C' \over 2} \left(\frac{A}{L}\right)^2 \Delta k
        \sum_{j=-j_{\rm mix}}^{j_{\rm mix}} \matrixelCl{q,p}{\MCl^j}{q,p}
    .
    \label{fourthmom}
\end{equation}

Comparing Eq.~(\ref{secondmom}) and Eq.~(\ref{fourthmom}) we obtain
for the LIPR
\begin{eqnarray}
    \text{LIPR} &=&
        \frac{N \sum_{n} \abs{\braket{q,p}{n}}^4}
        {\left(\sum_{n} \abs{\braket{q,p}{n}}^2\right)^2} \\
    &=& \frac{C'N}{2\Delta k}
        \frac{\sum_{j=-j_{\rm mix}}^{j_{\rm mix}}
            \matrixelCl{q,p}{\MCl^j}{q,p}}
        {\braketCl{q,p}{q,p \vphantom{\MCl^j} }}
    ,
\end{eqnarray}
where the projections in both the numerator and in the denominator are
classical.  The constant $C'$ can now be fixed by requiring that the
LIPR, when evaluated far from any short periodic orbits (where
scarring plays no role), should give the RMT value which we call
$\text{LIPR}(\text{RMT})$.  [As discussed above,
$\text{LIPR}(\text{RMT}) = 2$ in a system with no time-reversal
symmetry; in the presence of time-reversal symmetry,
$\text{LIPR}(\text{RMT})$ reaches the value of $3$ on certain symmetry
lines, as we have seen in Section~\ref{sec:rolesym}. The LIPR
contribution obtained from short-time dynamics must always be
multiplied by the appropriate factor given by symmetry considerations;
$\text{LIPR}(\text{RMT})$ is inherently an effect arising from
Heisenberg-time behavior and so must be considered separately from
short-time contributions to the LIPR.]

Away from short periodic orbits there will be no recurrences, so only
the $j = 0$ term in the above sum contributes.  Thus, $C' = (2 \Delta k/N) \,
\text{LIPR}(\text{RMT})$ and
\begin{equation}
    \text{LIPR}(q,p) = \text{LIPR}(\text{RMT})
        \frac{\sum_{j=-j_{\rm mix}}^{j_{\text{mix}}}
          \matrixelCl{q,p}{\MCl^j}{q,p}}
        {\braketCl{q,p}{q,p \vphantom{\MCl^j} }}.
    \label{eq:LIPR-dynam}
\end{equation}
The $j\neq 0$ terms of the sum give an enhancement to the LIPR
whenever a point $(q,p)$ is mapped near itself by an iterate of the
classical Poincar\'e map; i.e., near periodic orbits.  In the next
Section we will describe how to evaluate the LIPR in the neighborhood
of a periodic orbit.

\subsubsection{Periodic orbits, linear theory}
\label{scarsection}

The dynamical description of the LIPR developed in the previous
Section provides a method of computing the LIPR classically, anywhere
in phase space, by computing overlaps of classical Gaussian
probability distributions with themselves iterated under the classical
Poincar\'e map $\MCl$.  We found that the LIPR is enhanced when the
short-time
overlap is large.  This of course is most dramatically the case in the
neighborhood of periodic orbits.  In this Section we show how to
compute the LIPR in the neighborhood of a periodic orbit, using only
the properties of that one orbit.  We also generalize to the case of
stable and unstable manifolds with arbitrary orientation.

Fig.~\ref{fig:LIPR} 
confirms the omnipresence of
scarring in the eigenstates of the stadium billiard, as we now argue
(the identification of peaks in Fig.~\ref{fig:LIPR} with specific
short periodic orbits will be made below in Sec.~\ref{sec:return-prob},
\ref{identification-of-peaks} in particular).
In the general theory of scarring, a wavepacket launched on a periodic
orbit has a local density of states with a linear or short-time
envelope (coming from dynamics linearized around the orbit) that
consists of an oscillating function of energy;
isolated homoclinic recurrences,
if present, will give rise to further fluctuation multiplying this
envelope~\cite{KH2}.  The concentration of the local density of states
in some energy regions and avoidance of others leads to an enhanced
LIPR in Eq.~(\ref{eq:LIPR-def}).  Thus the peaks in the LIPR plot in
Fig.~\ref{fig:LIPR} at the positions of short periodic orbits must be
attributed to scarring, even though for the most part the scars are
too weak to be visible in coordinate-space plots of individual
wavefunctions, such as in Fig.~\ref{fig:wavefunctions}.  The scars are
there in the sense of non-$\chi^2$ variation of wavefunction intensity
on these periodic orbits, as one scans through the eigenstates of the
system.  Since there are several periodic orbits with appreciable
scarring, any given eigenvalue is likely to lie near a maximum in the
linear envelope of at least some of them, and the corresponding
eigenstate will more likely than not be scarred on these orbits, and
antiscarred on the others.
Examination of the spectra corresponding to different short periodic
orbits reveals that, first, as one varies the periodic orbit the
position of the smooth envelopes in the local density of states varies so as
to cause any given energy to lie under the peaks of the envelopes for
several different orbits.  Secondly, we find that most of the
envelopes in the local density of states are filled in an egalitarian
way (for a discussion of egalitarian versus totalitarian filling of
the spectral envelopes, see Kaplan and Heller\cite{KH2}), which
increases the likelihood of at least weak scarring along that orbit
for each eigenstate with energy near the maxima of the envelope.
For typical scarred
wavefunctions, the scarring will not be visible in coordinate space,
even though the phase space intensity is greatly enhanced at the
periodic points, as as one can see in the Husimi plots.  Occasionally,
the scarring may be strong enough to be also visible in a
coordinate-space plot; an illustration of strong multiple scarring can
be found in the state at $k=105.608$, which is simultaneously scarred
along the bowtie and diamond orbits; cf.\ 
Fig.~\ref{fig:wavefunctions}(c).

In summary, we find a distribution of wavefunction intensities on a
periodic orbit which differs from the $\chi^2$ distribution that would
be expected on the basis of random-matrix theory, as illustrated in
Fig.~\ref{fig:intensity-dist}.  In Fig.~\ref{fig:intensity-dist}(a) we
show a histogram of wavefunction intensities measured using a Gaussian
centered on the bowtie orbit for all states lying between $k=100$ and
$k=150$.  Because the bowtie orbit lies on a symmetry line, the
random-matrix-theory prediction is a $\chi^2$ distribution in one
degree of freedom (where we have set 6\% of the intensities to zero,
corresponding to the bouncing-ball states), 
which fails to account for the large number of high-intensity and low-intensity
peaks in Fig.~\ref{fig:intensity-dist}(a).
In the presence of scarring, the $\chi^2$ distribution must be
modulated by the linear spectral envelope, which, for a Lyapunov
exponent of 2.29 for the bowtie orbit, varies from 0.4 to 2.3
periodically in momentum.  The period in energy is given by $2\pi/T$
where $T=2.6T_B$ is the period of the bowtie orbit.  As a result of
this modulation, there are many more very small and very large
intensities in Fig.~\ref{fig:intensity-dist}(a) than would be expected
from the naive Porter-Thomas (RMT) picture.  
As plotted on a logarithmic scale, the effect is more pronounced for the 
larger intensities, but even at low intensities we expect an
enhancement by about 20\%, which is what is indicated by the data 
between {$\langle I \rangle/500$} and {$\langle I \rangle/10$, where
the statistics are good; below {$\langle I \rangle/500$} the statistics
are too poor to distinguish reliably between the scar theory and random 
matrix theory predictions.
The amount of strong scarring observed is also quantitatively 
consistent with the predicted scarring
corrections to random-matrix theory: for instance, the number of
wavefunction intensities greater than 10 times the mean is found to
be 10 numerically, while the scar-corrected prediction is $13 \pm 3$
and the uncorrected random-matrix prediction is only $3 \pm 3$.  At a
generic point in phase space, not lying on a scar or on a symmetry
line, we expect a $\chi^2$ distribution in two degrees of freedom, and
Fig.~\ref{fig:intensity-dist}(b) shows the numerical distribution of
wavefunction intensities at a generic point in phase space, in
satisfactory agreement with the random-matrix-theory prediction for
intensities greater than {$\sim \langle I \rangle/10$}
(without scarring corrections, since there is no scarring at a generic
point in phase space, but with 6\% of the intensities set to zero,
namely those corresponding to bouncing-ball states). The bouncing-ball
states appear of course in the numerical data at intensities 
{$ I < \langle I \rangle /10 $}; the total area between the data
and the curve in this intensity region
indeed turns out to be about 6\%).  Note the smaller
number of very small and very large intensities as compared to the
scarring case, Fig.~\ref{fig:intensity-dist}(a).  The
2-degrees-of-freedom distribution also falls off much faster at large
intensities than the 1-degree-of-freedom prediction appropriate for a
symmetry line.

We now turn to the calculation of the heights of the scarring peaks in
Fig.~\ref{fig:LIPR} based on a linear expansion around periodic
orbits.  Consider the surface-of-section map in Birkhoff coordinates;
we will be interested in the short-time linearized dynamics in the
vicinity of a periodic orbit~\cite{H1}.  The stable and unstable
invariant manifolds, defined as the locus of points that approach the
periodic orbit under infinite iteration of the mapping forwards or
backwards in time, respectively, will in general have an arbitrary
orientation.  In the special case that the invariant manifolds are
aligned with the coordinate axes centered on the periodic orbit of
interest---the simplest possibility---we have linearized
equations of motion $q(t) = q(0) e^{\lambda
t}$ and $p(t) = p(0) e^{-\lambda t}$, where $q$ and $p$ are measured
relative to the periodic point.  An initial Gaussian corresponding to
a classical distribution of probability, $P_0(q,p) = C
e^{-q^2/\sigma^2 - \sigma^2 p^2}$ (where $C = (2/\pi)^{1/2}$ is a
normalization constant) will map after $n$ iterations to
\begin{equation}
    P_j(q,p)=C \exp\left[
        - e^{-2 \lambda j} q^2 / \sigma^2
        - e^{2 \lambda j} \sigma^2 p^2
    \right]
    .
\end{equation}
The overlap with the initial distribution will then be
\begin{equation}
    \int dq \; dp \; P_j(q,p) P_0(q,p)
    = \frac{1}{\cosh\lambda j}
    \, ,
\label{clascosh}
\end{equation}
assuming that we are far into the semiclassical regime, i.e. that
$k$ is large enough so that the Gaussian wavepacket is well contained
in the phase space region where the linearized equations of motion are valid.
We note as an aside that there is no natural way to choose the normalization
constant $C$ based on purely classical considerations, since classically
the return probability of a distribution is not a meaningful concept. Instead,
we are implicitly thinking of Eq.~(\ref{clascosh}) as representing
the square of a {\it semi}-classical return amplitude of
a Gaussian wavepacket, which is a well-defined physical quantity, and is
given by the square root of Eq.~(\ref{clascosh}), times some irrelevant phase
associated with the action (in units of $\hbar$) of the periodic orbit. This
fixes the normalization constant $C$ above.

Due to the reloading effect mentioned in the preceding Section (see also
{\cite{KH2}}) in which returning homoclinic
orbits contribute in phase, the mean return probability for long times at
this periodic orbit, and hence the LIPR, which equals a constant times
the long-time return probability, will be equal to LIPR(RMT) times
the short-time enhancement factor
\begin{equation}
    S_1 \equiv
        \sum_{j=-\infty}^{\infty} \frac{1}{\cosh\lambda j}
    \, .
\label{s1def}
\end{equation}
Strictly speaking, the sum in Eq.~\ref{s1def} should extend only over times
short compared to the mixing time, i.e. $|j|<\lambda^{-1} \log(kL)$; however
this cutoff is irrelevant in the semiclassical limit $kL \gg 1$ in which
we are working.

It is desirable to extend this result to the case when the invariant
manifolds are not perpendicular.  This is equivalent to considering
perpendicular invariant manifolds but with a non-circular Gaussian (as
one can go from one case to the other via a canonical transformation).
Let $x = (q,p)$ and $A$ be the quadratic form defining the initial
Gaussian centered on a periodic orbit,
\begin{equation}
    P_0(x)=C e^{-x^T A x}
    \, ,
\end{equation}
with $\det A = 1$.  We define $J$ to be the Jacobian of $\MCl$ at the
periodic orbit; then
\begin{equation}
    P_j(x)=C \exp\left[ -x^T (J^{-j})^T A J^{-j} x \right]
\end{equation}
and the overlap is
\begin{equation}
    \int dq \; dp \; P_j(q,p)P_0(q,p)
    = 2 \sqrt{ \frac{\det A}{\det[A + (J^{-j})^T A J^{-j}]} }
\end{equation}
(again, we see that using our normalization the overlap is unity at time
$j=0$).
Thus, a better prediction of the LIPR peak heights at the periodic
orbits---one which takes into account the local stable and unstable
directions at the periodic orbit, and not only the stability
exponent---is LIPR(RMT) multiplied by
\begin{equation}
    S_2 \equiv \sum_{j=-\infty}^{\infty}
    2 \sqrt{ \frac{\det A}{\det[A + (J^{-j})^T A J^{-j}]} }
    \, .
\end{equation}
Note that $S_2$ reduces to $S_1$ when the invariant manifolds are
perpendicular, and the Gaussian test state is oriented along these
manifolds, as an easy calculation shows. Of course, using a canonical
transformation we may easily see that for any orientation of the manifolds
there is an infinite one-parameter family of `optimal' Gaussian test states,
all of which display maximum possible scarring in accordance with 
Eq.~(\ref{s1def}).
Both $S_1$ and $S_2$ approach unity for long or highly
unstable orbits, when the instability exponent $\lambda$ becomes
large.

We have calculated $S_1$ and $S_2$ for the periodic orbits appearing
in Fig.~\ref{fig:dynamical-structures}, and compare these theoretical
predictions with the actual peak heights in
Table~\ref{table:peak-heights}.
\begin{table}
    \begin{tabular}{|c|ccccccc|}
        Orbit & Symmetry & Lyapunov exponent & $S_1$ & $S_2$
            & actual & height & classical \\
        & & $\lambda$ & &
            & $k=100$-$150$ & $k=200$-$225$  & \\
        \hline
        HB    & a  & 1.76 & 2.72 & 2.15 & 1.86 & 2.01 & 2.78 \\
        V$_1$ & b  & 2.02 & 2.40 & 2.36 & 1.91 & 1.93 & 2.36 \\
        V$_2$ & c  &      & 2.40 & 1.95 & 1.68 & 1.87 & 1.97 \\
        D$_1$ & a  & 2.06 & 2.37 & 1.97 & 1.89 & 1.80 & 2.03 \\
        D$_2$ & b  &      & 2.37 & 2.27 & 2.15 & 2.06 & 2.27 \\
        B$_1$ & d  & 2.29 & 2.18 & 2.06 & 1.86 & 1.87 & 2.02 \\
        B$_2$ & d  &      & 2.18 & 2.06 & 1.86 & 1.87 & 2.02 \\
        Z$_1$ & e  & 2.38 & 2.10 & 2.04 & 1.69 & 1.88 & 2.04 \\
        Z$_2$ & e  &      & 2.10 & 2.04 & 1.69 & 1.88 & 2.04 \\
        Z$_3$ & c  &      & 2.10 & 1.77 & 1.53 & 1.55 & 1.85 \\
        X$_1$ & df & 2.58 & 2.00 & 1.94 & 1.91 & 2.03 & 2.07 \\
        X$_2$ & df &      & 2.00 & 1.94 & 1.91 & 2.03 & 2.07 \\
        A$_1$ & b  & 2.60 & 1.98 & 1.86 & 1.53 & 1.55 & 1.85 \\
        A$_2$ & d  &      & 1.98 & 1.86 & 1.59 & 1.51 & 1.90 \\
        A$_3$ & d  &      & 1.98 & 1.86 & 1.59 & 1.51 & 1.90 \\
        W$_1$ & b  & 2.68 & 1.94 & 1.94 & 1.71 & 1.86 & 1.95 \\
        W$_2$ &    &      & 1.29 & 1.27 & 1.68 & 1.78 & 1.29 \\
        W$_3$ & c  &      & 1.94 & 1.70 & 1.66 & 1.81 & 1.84 \\
        W$_4$ &    &      & 1.29 & 1.27 & 1.68 & 1.78 & 1.29 \\
        H$_1$ & a  & 3.26 & 1.74 & 1.73 & 2.01 & 2.37 & 2.00 \\
        H$_2$ & f  &      & 1.74 & 1.74 & 1.78 & 2.12 & 1.99 \\
        H$_3$ & f  &      & 1.74 & 1.73 & 1.78 & 2.12 & 2.00 \\
        T$_1$ & b  & 3.83 & 1.64 & 1.64 & 1.74 & 1.89 & 1.77 \\
        T$_2$ & f  &      & 1.64 & 1.56 & 1.15 & 1.26 & 1.61 \\
        T$_3$ & f  &      & 1.64 & 1.56 & 1.15 & 1.26 & 1.61 \\
        F$_1$ & a  & 4.90 & 1.55 & 1.53 & 1.58 & 1.63 & 1.95 \\
        F$_2$ &    &      & 1.03 & 1.04 & 1.14 & 1.18 & 1.02 \\
        F$_3$ & d  &      & 1.55 & 1.53 & 1.50 & 1.62 & 1.57 \\
        F$_4$ & d  &      & 1.55 & 1.55 & 1.50 & 1.62 & 1.57 \\
        F$_5$ &    &      & 1.03 & 1.03 & 1.14 & 1.18 & 1.02 \\
    \end{tabular}
    \vspace{3.0ex}
    \caption{Scar peak heights (values of LIPR/2) on short periodic orbits,
      as predicted by the linearized
      semiclassical measures $S_1$, $S_2$ and by the exact brute-force
      classical simulation, defined in the text, and to be compared with actual
      quantum-mechanically computed heights.
      Periodic orbit labels
      are defined in Table~\ref{table:orbits}. The symmetry family to
      which each point on each periodic orbit belongs is labeled as
      follows: $a$, orbit hits center of the circular endcap; $b$,
      hits center of the straight segment; $c$, incident normal to the
      boundary; $d$, possesses a vertical segment in the endcap
      region; $e$, passes through the center of the stadium; and $f$,
      possesses a horizontal segment.  The Lyapunov exponent is listed
      only once for each orbit. Again, the predictions $S_1$ and $S_2$
      are based on the tangent map in the desymmetrized stadium.  The
      predictions in columns 3, 4 and 7 include the quantum symmetry
      correction factor of $3/2$ for periodic points with special
      symmetry, as indicated in column 1.}
    \label{table:peak-heights}
\end{table}
All quantities appearing in Table~\ref{table:peak-heights} are either
predicted or actual values of ${\rm LIPR}(q,p)/2$ at a periodic point $(q,p)$.
In other words, we have divided out by the RMT prediction ${\rm LIPR}=2$, which
is relevant for all of space space away from the symmetry lines. However,
on the symmetry lines the RMT expectation is $3$ rather than $2$ 
(see Section~\ref{sec:rolesym}), so the theoretical values
$S_1$ and $S_2$ have been multiplied by $3/2$ for those
periodic points which do lie on symmetry lines,
in order to compare with the actual LIPR/2 data in the fifth and sixth
columns of Table~\ref{table:peak-heights}.
In fact,
most of the points listed in Table~\ref{table:peak-heights} do lie on
one or another symmetry line; the first column of the table indicates
which symmetry class each point lies in (see definition of the symbols
in the caption).  As expected, $S_1$ fails to be an accurate measure
of the LIPR peak heights because the test states in general are not
aligned optimally with the invariant manifolds.  With the inclusion of
symmetry corrections, the agreement of $S_2$ is roughly correct, as
can be seen in a scatter plot of the predicted versus calculated LIPR/2
peak intensities, Fig.~\ref{fig:scatter-plots}; as shown in
Fig.~\ref{fig:scatter-plots}(c), the scatter is markedly worse when
the symmetry corrections are omitted.

In order to appreciate the significance of these results, we should
discuss the factors contributing to the uncertainty in the measured
LIPR/2 values
listed in columns 6 and 7 of Table~\ref{table:peak-heights}.
There are two issues: first, there may be a non-integral number of
oscillations of the local density of states in the energy window 
considered.  The size of this effect scales inversely with the number
of scar oscillations in the energy window, and is only a few percent
for the data presented in Table~\ref{table:peak-heights}.  A second,
and bigger, source of error arises from the finite number of
eigenvalues in the window.  For random fluctuations, the uncertainty
coming from this effect scales as the square root of the mean level
spacing divided by the size of the window.  This is why the windows
from $k=100$ to $150$ and from $k=200$ to $225$ were taken to be as
large as possible.  A monte carlo simulation was done of random
wavefunction intensities constrained by the ``linear'' spectral
envelope: it suggests that the statistical uncertainty in the LIPRs
should range from $5\%$ to $8\%$ (depending on the stability matrix of
the orbit in question) for periodic points on symmetry lines, and
should be slightly smaller for those points that are not on any
symmetry line.  This level of fluctuation is roughly consistent with the
dispersion in the scatter plots of Fig.~\ref{fig:scatter-plots}, and
does not alter the conclusion that $S_2$ together with symmetry
factors is a good prediction of the peak heights.  Thus, the symmetry
lines emerge as essential to a quantitative understanding of the
calculated LIPRs due to scarring.  A final consideration supporting the
validity of our symmetry analysis is the pattern of peak heights for
the F orbit, on which periodic points $\text{F}_1$, $\text{F}_3$ and
$\text{F}_4$ have a symmetry correction while $\text{F}_2$ and
$\text{F}_5$ do not; thus, the predicted peak heights for the points
having symmetry corrections are about fifty percent higher than for
the others, and this predicted pattern is reproduced in the calculated
heights.

The brute-force classical calculation at times large compared to $T_B$
goes beyond $S_1$ and $S_2$ to include the effects of nonlinear
homoclinic recurrences, which add to the predicted peak height.  For
each periodic point, the returning probability for an initial Gaussian
distribution centered at that point was integrated up to a cutoff time
of $5 T_B$, which was chosen to be large compared to $T_B$ but smaller
than the mixing time $T_{\rm mix} \approx 7 T_B$, after which the
returning probability approaches a constant per unit time independent
of position in phase space.  In the semiclassical limit, where
strong, identifiable
recurrences at times between the initial decay of the wavepacket and the 
Heisenberg time can be neglected, the classical simulation
reproduces $S_2$. 
The classical simulation
reproduces, as it must, one feature of the quantum data: the peak
heights it predicts agree for points related by symmetry, namely
$\text{B}_1$ and $\text{B}_2$, $\text{Z}_1$ and $\text{Z}_2$,
$\text{X}_1$ and $\text{X}_2$, $\text{A}_1$ and $\text{A}_2$,
$\text{W}_2$ and $\text{W}_4$, $\text{H}_2$ and $\text{H}_3$,
$\text{T}_2$ and $\text{T}_3$ and $\text{F}_3$ and $\text{F}_4$.  In
predicting the quantum heights, the brute-force classical calculation
does no better than $S_2$ and makes predictions that are
systematically high (after including symmetry effects).  In order to
investigate the energy dependence of the classically predicted peak
heights, the classical simulation was run both at $k=100$ and at
$k=200$, the difference being in the size of the Gaussian as a
fraction of the phase space area.  For almost all of the periodic
orbits the results at the two energies agreed to within a few percent,
which is of the same size as the discrepancy between the numerical
quantum peak heights in those two energy ranges. This tells us that
isolated, identifiable non-linear recurrences beyond the initial decay
of the wavepacket away from the periodic orbit are not that important for
computing the mean long-time return probability.
Any such nonlinear effects would be strongly dependent on the size of the 
initial wavepacket, i.e. on the wavevector $k$.

The classical calculation for the average return probability can be
performed anywhere in phase space, not only on the short periodic orbits
considered in this Section; see Sec.~\ref{class-return-prob} below.
There we shall see that we get a picture similar to Fig.~\ref{fig:LIPR},
but with more diffuse peaks and the onset of mixing for {$T>T_{\rm mix}$}.

\subsection{Dynamics plus symmetry}
\label{sec:dynamsym}

In the previous two Sections, we have seen that 
discrete symmetries
have two important and distinct
effects on wavefunction localization properties. First,
in the presence of time-reversal symmetry, the horizontal
line of zero parallel
momentum in the surface of section phase space, as well as vertical
lines associated with parity symmetry, if any, both have a local IPR
of $3$, in contrast with the background LIPR of $2$ that is seen away from
the symmetry lines. The values LIPR$\;=3$ and LIPR$\;=2$ correspond to the
random matrix theory predictions for real and complex Gaussian random
wavefunctions, respectively. Furthermore, short-time iterates of these
symmetry lines result in additional (curved) symmetry lines where the LIPR
takes values intermediate between $2$ and $3$. When a periodic point happens
to be located on a symmetry line (which in a highly symmetric system such as 
the stadium happens quite often for the short orbits), the LIPR enhancement 
associated with scar effects must be combined with the enhancement due
to symmetry. In the semiclassical limit where our statistical analysis is 
relevant, the two effects are independent of one another and simply multiply,
the scar effect being associated with short-time dynamics around a periodic
orbit, while the symmetry effect appears near the much larger Heisenberg time,
where individual eigenlevels are resolved.

Secondly, in the presence of spatial symmetries such as parity,
symmetry must be taken into
consideration when computing periodic orbit properties (period,
classical action and monodromy matrix) for the purpose of quantifying the
scar effect. This holds true even if the system is not time-reversal
invariant, so that no symmetry lines are present in the LIPR plot. In the
previous Section, we have addressed this second issue simply by considering 
orbits in a desymmetrized version of the billiard. That approach, however,
is neither rigorous nor completely general, and in particular is problematic
for orbits like the horizontal bounce, which does not exist in the interior
of the quarter-stadium. In this Section we present an alternative approach to
including spatial symmetry in computing scar effects on localization.

In the stadium there are two parity
operations, reflection in {$x$} and in {$y$}, which we represent
by the operators {$R_x$}
and {$R_y$}, respectively, and which separate the stadium wavefunctions
into four distinct symmetry classes.
The sums on the left-hand side in Eqs.~(\ref{secondmom}) and (\ref{fourthmom}) 
can be divided into sums over each symmetry class separately; thus, for 
instance, the left-hand side of Eq.~(\ref{secondmom}) becomes
\begin{eqnarray}
    \sum_n \abs{\braket{q,p}{n}}^2 & = &
        \sum_{n_{++}} \abs{\braket{q,p}{n_{++}}}^2 +
        \sum_{n_{-+}} \abs{\braket{q,p}{n_{-+}}}^2 +
        \sum_{n_{+-}} \abs{\braket{q,p}{n_{+-}}}^2 +
        \sum_{n_{--}} \abs{\braket{q,p}{n_{--}}}^2 \\
    &=& 
    \sum_n \left[
        \abs{\matrixel{q,p}{P_{++}}{n}}^2 +
        \abs{\matrixel{q,p}{P_{-+}}{n}}^2 +
        \abs{\matrixel{q,p}{P_{+-}}{n}}^2 +
        \abs{\matrixel{q,p}{P_{--}}{n}}^2 \right],
\end{eqnarray}
where {$P_{\pm\pm}$} is the projector onto the subspace with even or
odd symmetry under {$R_x$} and {$R_y$}, respectively:
{$P_{\pm\pm}=(1\pm R_x)(1\pm R_y)/4$}.
A similar development of the left-hand side of Eq.~(\ref{fourthmom}) for
the fourth moment is possible.  The projection operators may now be considered
to act to their left, on the state {$|p,q\rangle$}.  This means simply that
{$|p,q\rangle$} is to be replaced by {$P_{++}|p,q\rangle$}, etc., on 
the right-hand sides of Eqs.~(\ref{secondmom}) and (\ref{fourthmom}).
If we consider the LIPR for the even-even states only we obtain
\begin{eqnarray}
    \text{LIPR}_{++}(q,p) & = &
        \frac{N_{++} \sum_{n} \abs{\matrixel{q,p}{P_{++}}{n}}^4}
        {\left( \sum_n \abs{\matrixel{q,p}{P_{++}}{n}}^2 \right)^2} \\
    & \approx & 
        {\rm LIPR(RMT)}
          {{\sum_{j=-j_{\rm mix}}^{j_{\rm mix}}
          \left[ \matrixel{q,p}{P_{++} M^jP_{++}}{q,p}\right]^2} \over
        {|\braket{q,p|P_{++}}{q,p}|^2}} \\
    & = & {\rm LIPR(RMT)} \sum_{j=-j_{\rm mix}}^{j_{\rm mix}}
        \frac{\left[ \frac{1}{4} \sum_{i_1,i_2=0,1}
              \matrixel{q,p}{R_x^{i_1}R_y^{i_2} M^j
                }{q,p}\right]^2}
        {|\braket{q,p|P_{++}}{q,p}|^2} \\
    & \approx & {\rm LIPR(RMT)} \sum_{j=-j_{\rm mix}}^{j_{\rm mix}}
          \frac{\left[ \frac{1}{4} \sum_{i_1,i_2=0,1}
          \matrixelCl{q,p}{R_x^{i_1}R_y^{i_2}\MCl^j}{q,p}^{1/2}\right]^2}
          {\braketCl{q,p|P_{++}}{q,p}}
    \label{eq:iprrecur}
    .
\end{eqnarray}
The constant of proportionality here, {$C'_{++}=(2 \Delta k / N_{++})
{\rm LIPR}({\rm RMT})$}, is four times larger than {$C'$} in
Eq.~(\ref{fourthmomq})
because
{$N_{++}=N/4$}; this reflects the fact that the Heisenberg time is
four times smaller for the desymmetrized stadium ({$C'_{++}$} being
inversely proportional to {$T_{H++}$}).

Thus we see that in addition to the usual recurrences due to periodic
orbits ($M^j(q,p) = (q,p), i_1, i_2 = 0$) there are, for
certain initial values $(q,p)$, recurrences when the orbit closes up
to a reflection in $x$ or $y$ or both.  For example, the V
orbit starts at the center of the straight segment, hits the
semicircular segment at $j = 1$, and at $j = 2$ returns to where it
started up to reflection in $x$.  Another example is the Z orbit, which
passes through the center of the stadium; at $j = 3$ it returns to
where it started up to a reflection in both $x$ and $y$.

Similar recurrences up to a symmetry operation would be present in the LIPR
summed over all symmetry classes. One last important point to note is that
for an orbit located right on a symmetry line (such as the horizontal bounce
orbit in the stadium), the corresponding symmetry ($R_y$ in this case) is
not included in computing the scar effect. The fact that the wavepacket is
located on a symmetry line is of course relevant in setting 
LIPR(RMT)$=3$ instead of $2$. The other symmetry, namely $R_x$, still needs
to be included: for the horizontal bounce orbit its effect
is to make short-time
recurrences in Eq.~\ref{eq:iprrecur} happen for all values of $j$
instead of for even $j$ only.

\subsection{Derivation as return probability}
\label{sec:return-prob}

Finally, we explain the interpretation of the LIPR as the
infinite-time average return probability, as in Eq.~(\ref{eq:2}) below
(where it is assumed that the energy window in the sums over $n$ and
$n'$ is large enough to cover the energy scales present in the test
state $\ket{a}$).  In order to clarify what is happening in the
infinite-time limit it is useful to look at finite times, and then let
time tend to infinity.  Thus we define the average return probability
at time $T$ for a wavepacket $\ket{\Psi_a}$ to be~\cite{old}
\begin{eqnarray}
    P_{aa}(T) & = & \recip{2T}
    \int_{-T}^T dt \abs{ \matrixel{\Psi_a}{e^{-iHt}}{\Psi_a} }^2
    \nonumber \\
    & = & \recip{2T} \int_{-T}^T dt
        \abs{ \sum_n \braket{\Psi_a}{n} e^{-i E_n t} \braket{n}{\Psi_a} }^2
    \nonumber \\
    & = & \recip{2T} \sum_{n n'} \int_{-T}^T dt
        \abs{\braket{\Psi_a}{n}}^2 \abs{\braket{\Psi_a}{n'}}^2
        e^{-i(E_n - E_{n'})t}
    \nonumber \\
    & = & \sum_{n n'}
        \frac{\sin\left((E_n-E_{n'})T\right)}{(E_n-E_{n'})T}
        \abs{\braket{\Psi_a}{n}}^2 \abs{\braket{\Psi_a}{n'}}^2 \,.
    \label{eq:2}
\end{eqnarray}
In the limit as $T$ tends to infinity,
\[
    \frac{\sin\left( (E_n - E_{n'})T \right)}{(E_n-E_{n'})T}
    \rightarrow \delta(E_n - E_{n'}) 
\]
and (in the case of a non-degenerate spectrum) we recover
Eq.~(\ref{eq:LIPR-def}) up to an overall proportionality
constant. The
arbitrary state $\ket{\Psi_a}$ living in the interior of the stadium
may be replaced by a Gaussian $\ket{q,p}$ living on the boundary if at
the same time the eigenstates $\ket{\Psi_n}$ are replaced by their
normal derivatives on the boundary, $\ket{n}$.
The natural timescale for $T$ is the time $T_B$ for bouncing between
the straight segments of the billiard.  With $k = 100$ and $m = 1/2$
we have $v = p/m = \hbar k/m = 200$ and so $T_B = 2/v = 0.01$.
For $T \ll T_B$ we should have $P_{aa}(T) \sim 1/T$; for intermediate
times we expect classical behavior as long as $T < T_{\rm mix}$, where
the mixing time $T_{\text{mix}}$ is the time it takes for the classical
dynamics to spread a given phase-space element throughout the entire
phase space (in order of magnitude it is given by $T_B$ times the
logarithm of the number of Planck-sized cells in the classical
surface-of-section
phase space, or $\sim T_B \log(kL)$  ).
The classical behavior will be computed directly below, where based on
the time needed for a simulated classical ensemble to begin to spread
smoothly throughout phase space we find $T_{\rm mix} \approx 7 T_B$ at
$k = 100$; that it is so large may be attributed to intermittency due
to bouncing-ball orbits.  Finally for $T \gg T_H$, that is beyond the
Heisenberg time where individual eigenstates are resolved (here, $T_H
= 1/\Delta E = 1/(2k\Delta k) = 1/(2 \cdot 100 \cdot 0.05) = 0.1 = 10
T_B$ at $k=100$), we should find that $P_{aa}(T)$ tends to the LIPR plot
as shown in Fig.~\ref{fig:LIPR}.

\subsubsection{Finite-time return probability}

The results for the time-dependent average return probability
$P_{(q,p),(q,p)}(T)$ are given in Fig.~\ref{fig:Paa(T)}, for the
energy window at $k=100$.  We are unable to study numerically the
limit $T \ll T_B$ because of the finite number of eigenfunctions
included in the sum in Eq.~(\ref{eq:2}).  At $T= T_B$, however, we see
large amplitudes coming from the bouncing-ball and whispering-gallery
regions of phase space, both of which are short-time classical
effects.  At $T = 2 T_B$ the horizontal-bounce orbit appears.  By $T =
3 T_B$ several discrete peaks appear, and more of them are present at
$T = 4 T_B$.
These will be shown below to correspond to short
periodic orbits of the classical dynamics in the desymmetrized
quarter-stadium billiard.  As $T$ becomes as large as $5 T_B$ and then
$10 T_B$ the finite-time average return probability indeed approaches
the infinite-time LIPR.  The non-uniform structure of the
infinite-time return probability is a purely quantum-mechanical
effect, because according to classical mechanics the average return
probability must become uniform for $T \gg T_{\text{mix}}$.

\subsubsection{Identification of peaks}
\label{identification-of-peaks}

Let us work with the data at $T=4T_B$, where the strong peaks are all
present and readily distinguishable from the background [see
Fig.~\ref{fig:Paa(T)}(d)].  We identify these peaks with the short
periodic orbits in Table~\ref{table:orbits} and
Fig.~\ref{fig:dynamical-structures}.
\begin{table}
    \begin{tabular}{clcccc}
        Label & Description & $q$ & $p/k$ & Length & $\text{Period}/T_B$ \\
        &             &     &     & (desymmetrized) & (desymmetrized) \\
        \hline
        HB & horizontal bounce & 2.5708 & 0       & 1 & 2.00 \\
        V  & V-shaped          & 0      & 0.70711 & 2 & 2.41 \\
        D  & diamond           & 2.5708 & 0.44721 & 2 & 2.24 \\
        B  & bowtie            & 1.5236 & 0.5     & 2 & 2.60 \\
        Z  & Z-shaped          & 0.5    &-0.44721 & 3 & 3.24 \\
        X  & box               & 1.7854 & 0.70711 & 2 & 2.41 \\
        A  & accordion         & 0      & 0.5491  & 3 & 3.30 \\
        W  & W-shaped          & 0      & 0.31623 & 4 & 4.16 \\
        H  & hexagon           & 2.5708 &-0.86603 & 3 & 2.50 \\
        T  & triangle          & 0      & 0.83573 & 3 & 4.30 \\
        F  & fish              & 2.5708 & 0.6478  & 5 & 5.40 \\
    \end{tabular}
    \vspace{3.0ex}
    \caption{%
      Principal short periodic orbits appearing in
      Figs.~\ref{fig:LIPR} and \ref{fig:Paa(T)}.  The initial
      coordinate $q$ is the distance around the perimeter of the
      stadium measured from the center of the upper straight segment;
      the initial momentum $p$ is positive in the clockwise sense.
      The lengths and periods are given for the desymmetrized stadium,
      which is relevant to our calculations.}
    \label{table:orbits}
\end{table}
To avoid having to deal explicitly with
symmetry effects, we use the properties of the reduced orbits
in the fundamental domain---one quarter of the stadium.  (This
approach has been justified rigorously in Sec.~\ref{sec:dynamsym}.)
Thus the periods of the most important orbits, as tabulated in
Table~\ref{table:orbits}, range from $2T_B$ to $5T_B$ although their
periods in the full stadium may be longer.  Although the triangle is a
short periodic orbit, it does not appear until $5T_B$ because its
desymmetrized version is just as long as the orbit in the full
stadium.
As $T$ increases beyond $T_H \approx 10 T_B$, many of the short
periodic orbits become incorporated into the streaks that dominate the
IPR plots.

\subsubsection{Classical finite-time return probability}
\label{class-return-prob}

The persistence to long times of short-time classical information is a
remarkable property of the quantum mechanics.  Classically, we would
expect the recurrences to be washed out at times large compared to the
mixing time.  Indeed, we have computed a classical analogue to the
time-dependent average return probability, namely the overlap of an
initial Gaussian in classical (Birkhoff) phase space with its iterates
under the classical mapping.  In order to get results in continuous
time, the overlap of the iterate at (real) time $t$ with the initial
Gaussian was stored as a delta function at time $t$ and at the end of
the calculation the integral over $t$ from $-T$ to $+T$ was taken, as
in the quantum calculation.  As shown in
Fig.~\ref{fig:LIPR-classical}, by $T = 10 T_B$ all of the periodic
orbits seen in the quantum calculation are present, although the peaks
are more diffuse classically, but by $20 T_B$ the classical picture is
beginning to be uniformly distributed throughout phase space, losing
memory of the short-time behavior.
Quantum-mechanically, of course, the peaks do not become washed out at 
long times and persist in the infinite-time limit, Fig.~\ref{fig:LIPR}.

\section{Conclusion}
\label{conc}

The advantage of the approach implemented in Section \ref{locdens} is
that all of the scars can be observed simultaneously in the plot of
inverse participation ratio as a function of phase space location.
Although scars are associated with the classical structure of unstable
periodic orbits, the persistence to long times of an excess in the
average return probability is a purely quantum-mechanical phenomenon.
The brightnesses of the peaks in the LIPR plot can be predicted
semiclassically with some success (to accuracy of about $10 - 15\%$),
the size of the deviations being roughly consistent with expected
statistical
fluctuations.  More precise comparison with the scar theory predictions
would require producing a greatly increased set of eigenstates: this
could be done by going to higher energies or (with far less
computational expense) by using an ensemble of chaotic systems with a
given unstable orbit kept fixed~\cite{KH2}. The predictions of
brightnesses based on exact classical phase-space evolution at the end
of Section~\ref{locdens} were no more successful than those based on
linearized behavior around the periodic orbit.

Most of the extra localization of individual eigenfunctions in Husimi phase
space, over and above Gaussian random fluctuations, may be
satisfactorily accounted for by quantum symmetry effects.  Scarring
is also seen to be present based on an analysis of the spectra of
wavepackets located on short periodic orbits.  Symmetry effects on the
overall mean IPR scale as $O(\sqrt{\hbar})$ (i.e., $O(\sqrt{1/k})$),
while the effect of scarring on
the overall mean IPR
scales as $O(\hbar)$ (i.e., $O(1/k)$).
The reason for this, for the symmetry effects,
is that the widths in position and momentum of the test-state Gaussian
scale as $\sigma \sim \sqrt{\hbar}$, setting the width of the symmetry
lines (Eqs.~(\ref{eq:sym-q=0}),~(\ref{eq:sym-p=0})), while the length
of the symmetry-enhanced lines is $\hbar$-independent.  Thus, the total
fraction of phase space covered by the symmetry lines scales as
$\sqrt{\hbar}$.  The scar effect, on the other hand, is significant
for those phase-space Gaussian wavepackets that have large intensity
on the periodic orbit~\cite{H2}, and the fraction of such Gaussians
goes as the area of the Gaussian; i.e., as $\sqrt\hbar \times
\sqrt\hbar = \hbar$. We note that although the size of both these effects
on the overall wavefunction IPR is expected to go to zero in the semiclassical
limit (because periodic orbits and symmetry lines both affect a smaller and
small fraction of phase space surrounding them, in the $\hbar \to 0$ limit),
the local IPR (LIPR) either on a symmetry line or on a periodic orbit
is $\hbar$-independent. Thus, as measured using the local IPR, deviations
from naive RMT predictions persist to arbitrarily high energies and do
not decay in the semiclassical limit.
As for other possible kinds of eigenfunction
localization, not associated with the short periodic orbits of the
system or with symmetry lines, their explanation remains an open question.

\section{Acknowledgments}

This research was supported by the National Science Foundation under
Grant No.\ CHE-9610501.

\newpage

\begin{figure}

\centerline{
\psfig{file=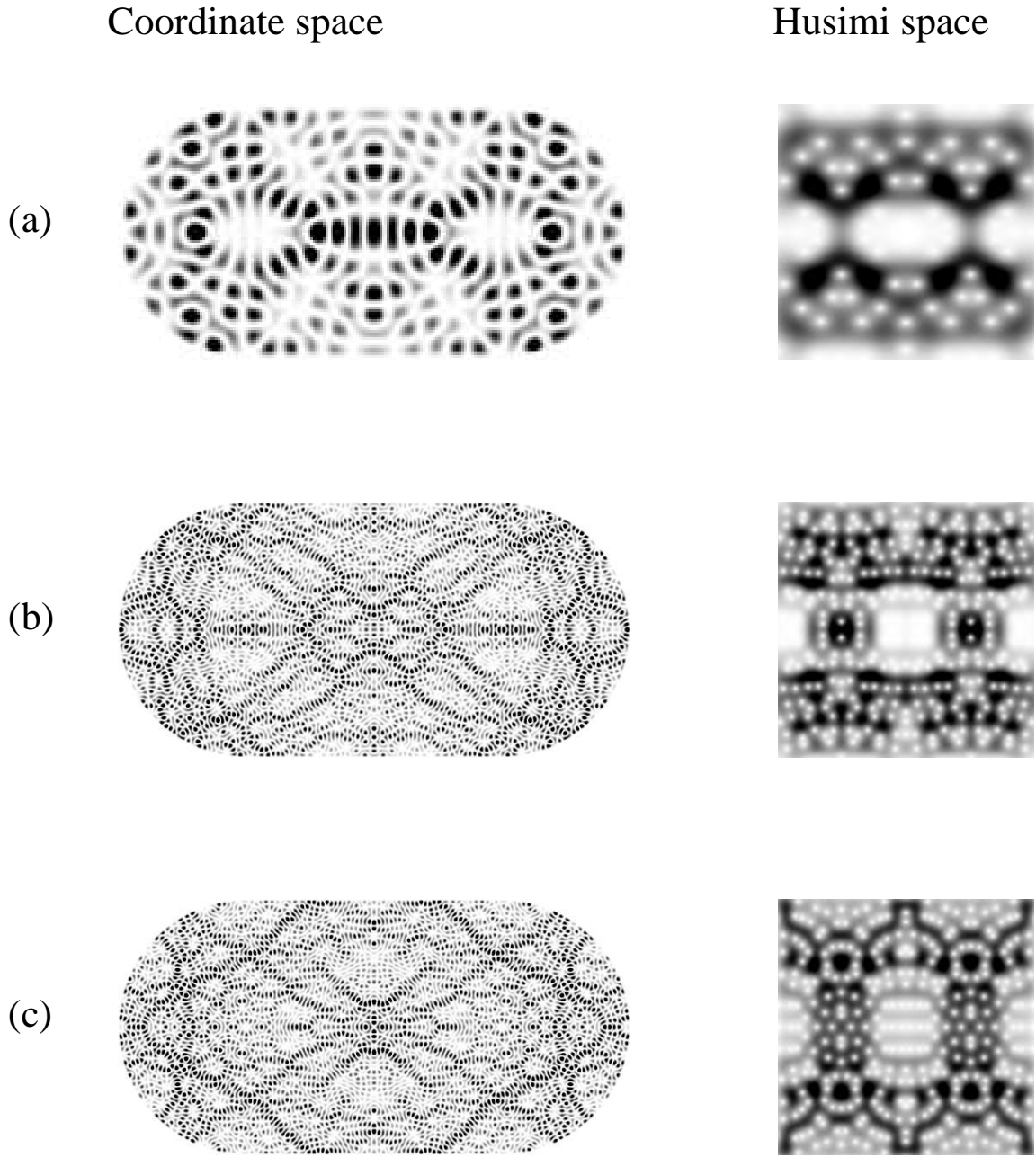,width=6in,bbllx=25pt,bblly=80pt,bburx=585pt,bbury=715pt,clip=}}

\vskip 0.2in

\caption{Representative eigenfunctions of the stadium billiard in
  coordinate space (left) with $x$-axis running from 0 to 4 and
  $y$-axis running from $-1$ to 1, and in Husimi phase space (right) with
  distance $q$ along the perimeter (horizontal axis) running clockwise
  from $0$ (corresponding to the center of the upper straight segment)
  to $4+2\pi$ and
  tangential momentum (vertical axis) running from $-k$ to $k$, where
  $k$ is the wavenumber of the eigenfunction.  In the coordinate space
  plots, the graylevel represents {$|\psi|^2$}. In the Husimi phase 
  space plots, white is high
  intensity and black low intensity.  (a) $k=24.680$, (b) $k=100.787$
  and (c) $k=105.608$, chosen to illustrate scarring along the bowtie
  and diamond orbits (cf.\ Table~\ref{table:orbits}).
}

\newpage

\label{fig:wavefunctions}
\end{figure}

\newpage

\begin{figure}

\centerline{
\psfig{file=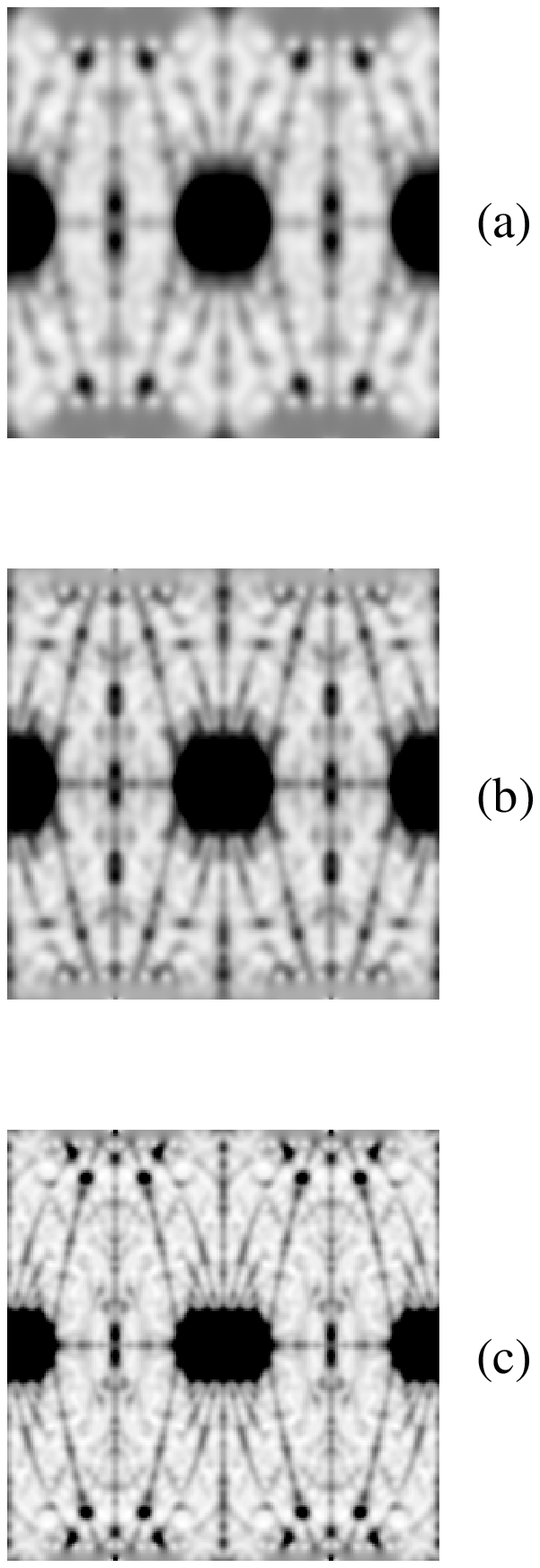,width=2.5in,bbllx=215pt,bblly=145pt,bburx=395pt,bbury=650pt,clip=}}

\vskip 0.1in
\caption{LIPR (see Eq.~\ref{eq:LIPR-def}) in phase space as a function of
  $q$ running horizontally from $0$ to $4+2\pi$ and $p/k$ running
  vertically from $-1$ to $1$.  The test-state width $\sigma$ is
  chosen so that the test Gaussians have circular cross-sections when
  the figures are plotted with a square aspect ratio.  Darker
  shading indicates higher values.  (a) $k=50$ to $60$, (b) $k=100$ to
  $150$, and (c) $k=200$ to $225$.}

\label{fig:LIPR}
\end{figure}

\newpage

\begin{figure}

\centerline{
\psfig{file=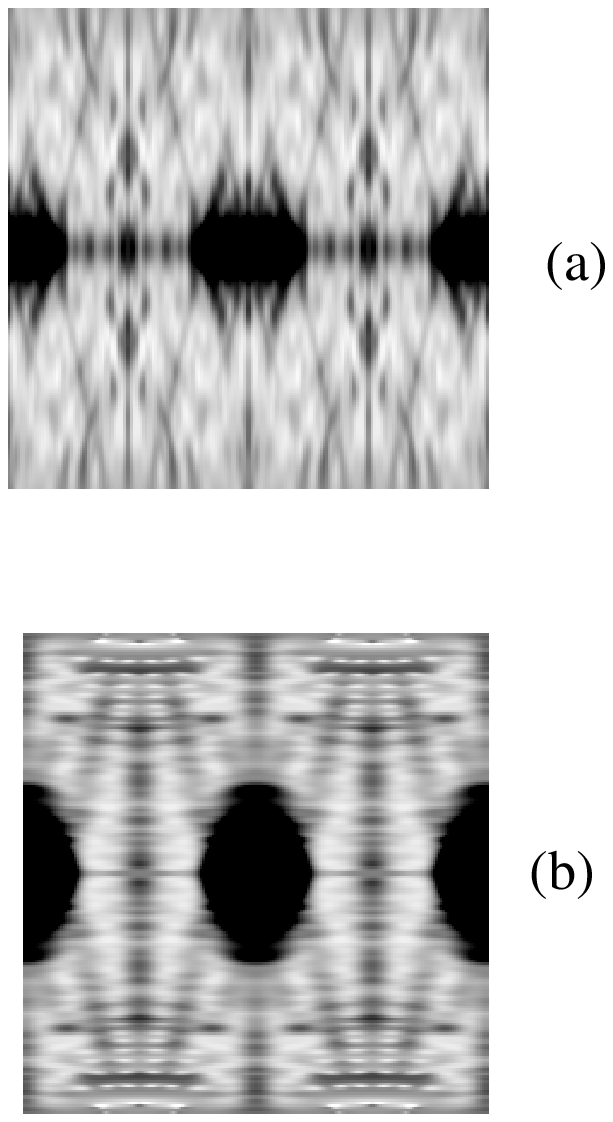,width=3in,bbllx=215pt,bblly=235pt,bburx=400pt,bbury=560pt,clip=}}

\vskip 0.1in
\caption{LIPR in phase space for eigenfunctions in the range from
  $k=100$ to $150$, coordinates as in Fig.~\ref{fig:LIPR}(b), for (a)
  $\sigma^2 \approx 0.01$, position-like test Gaussians, to be
  compared with $\sigma^2 \approx 0.05$ used in
  Fig.~\ref{fig:LIPR}(b), and (b) $\sigma^2 \approx 0.25$,
  momentum-like test Gaussians.}

\label{fig:LIPR-aspect}
\end{figure}

\newpage

\begin{figure}

\centerline{
\psfig{file=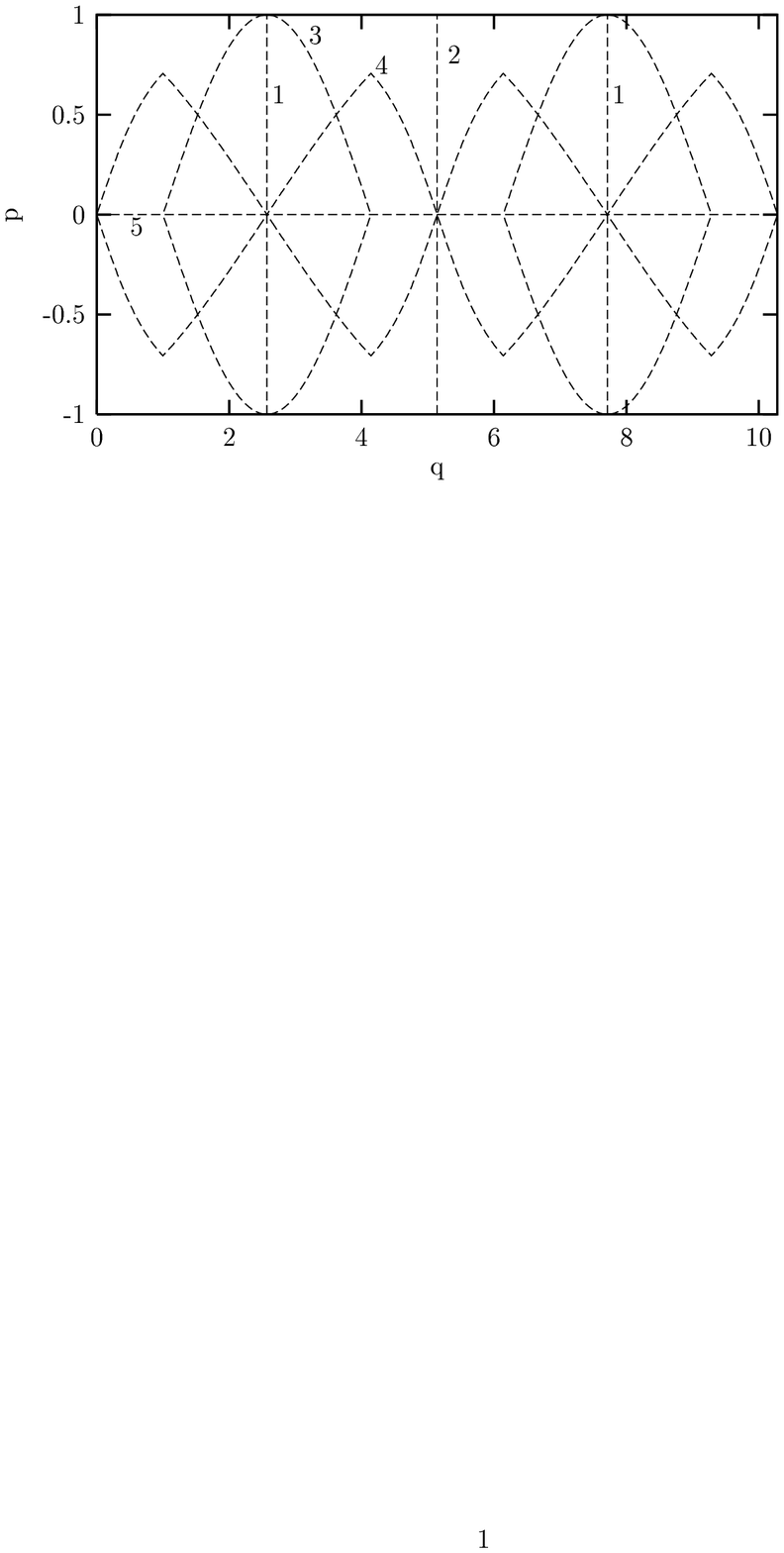,width=6in,bbllx=85pt,bblly=500pt,bburx=445pt,bbury=720pt,clip=}}

\vskip 0.2in

\caption{Principal symmetry-related structures in phase space of the
  inverse-participation ratio plotted in Fig.~\ref{fig:LIPR} and the
  average-return probability plotted in Fig.~\ref{fig:Paa(T)}.
  Schematic depiction of the symmetry lines in Fig.~\ref{fig:LIPR},
  labeled in the order in which they are described in the text.}

\newpage

\label{fig:symmetry-structures}
\end{figure}

\newpage

\begin{figure}

\centerline{
\psfig{file=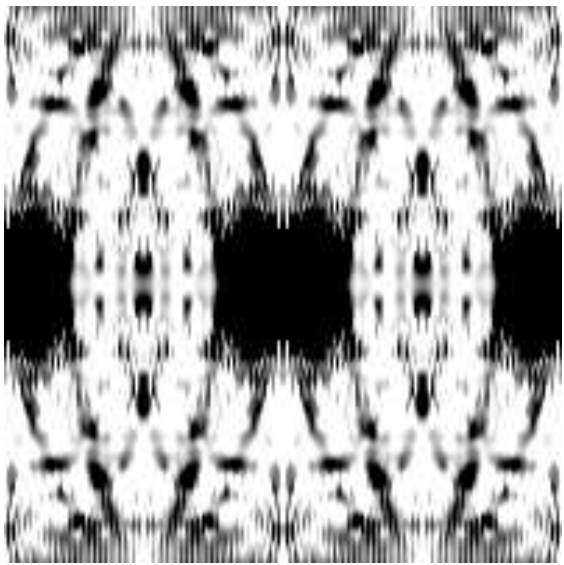,width=3in,bbllx=0pt,bblly=0pt,bburx=286pt,bbury=286pt,clip=}}

\vskip 0.1in
\caption{LIPR for real test wavepcakets
  (see Eq.~(\ref{eq:LIPR-desym-def})) in
  phase space as a function of $q$ running horizontally from $0$ to
  $4+2\pi$ and $p/k$ running vertically from $-1$ to $1$.
  The axes and parameters are the same as
  those used in Fig.~\ref{fig:LIPR}.}

\label{fig:LIPR-desym}
\end{figure}

\newpage

\begin{figure}
\centerline{
\psfig{file=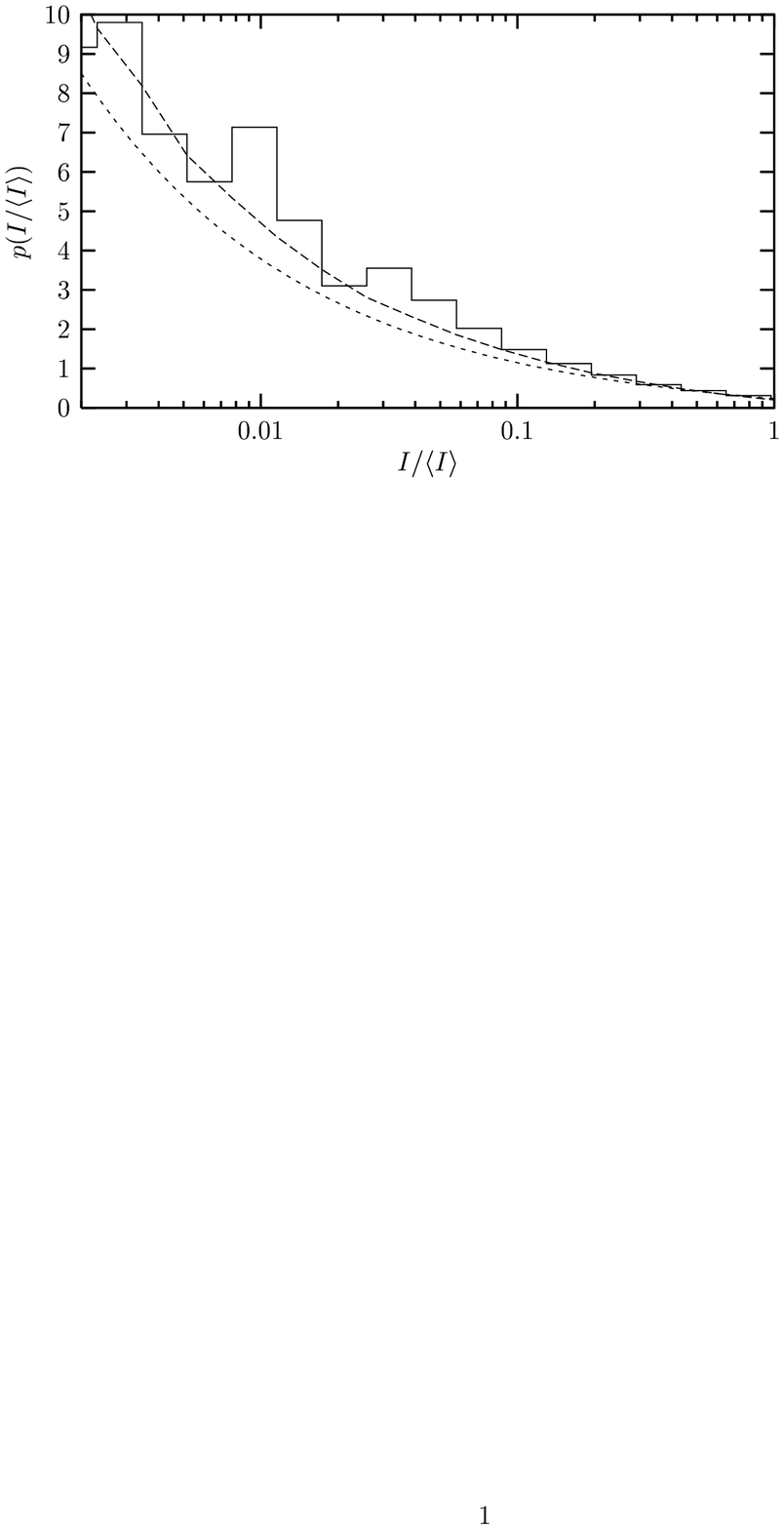,width=3in,bbllx=85pt,bblly=500pt,bburx=445pt,bbury=720pt,clip=}
\psfig{file=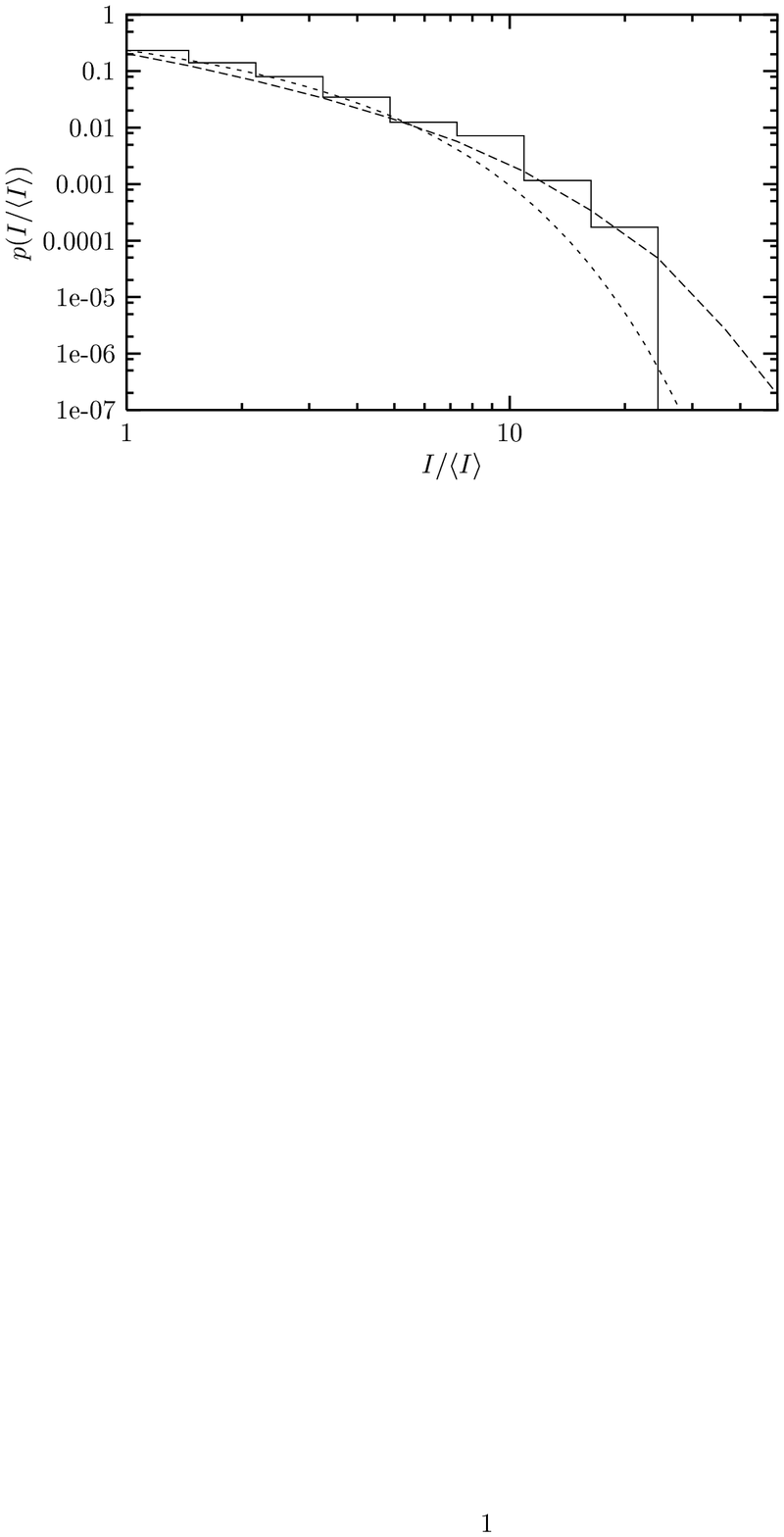,width=3in,bbllx=85pt,bblly=500pt,bburx=445pt,bbury=720pt,clip=}
}

\centerline{(a)}

\vskip 0.2in

\centerline{
\psfig{file=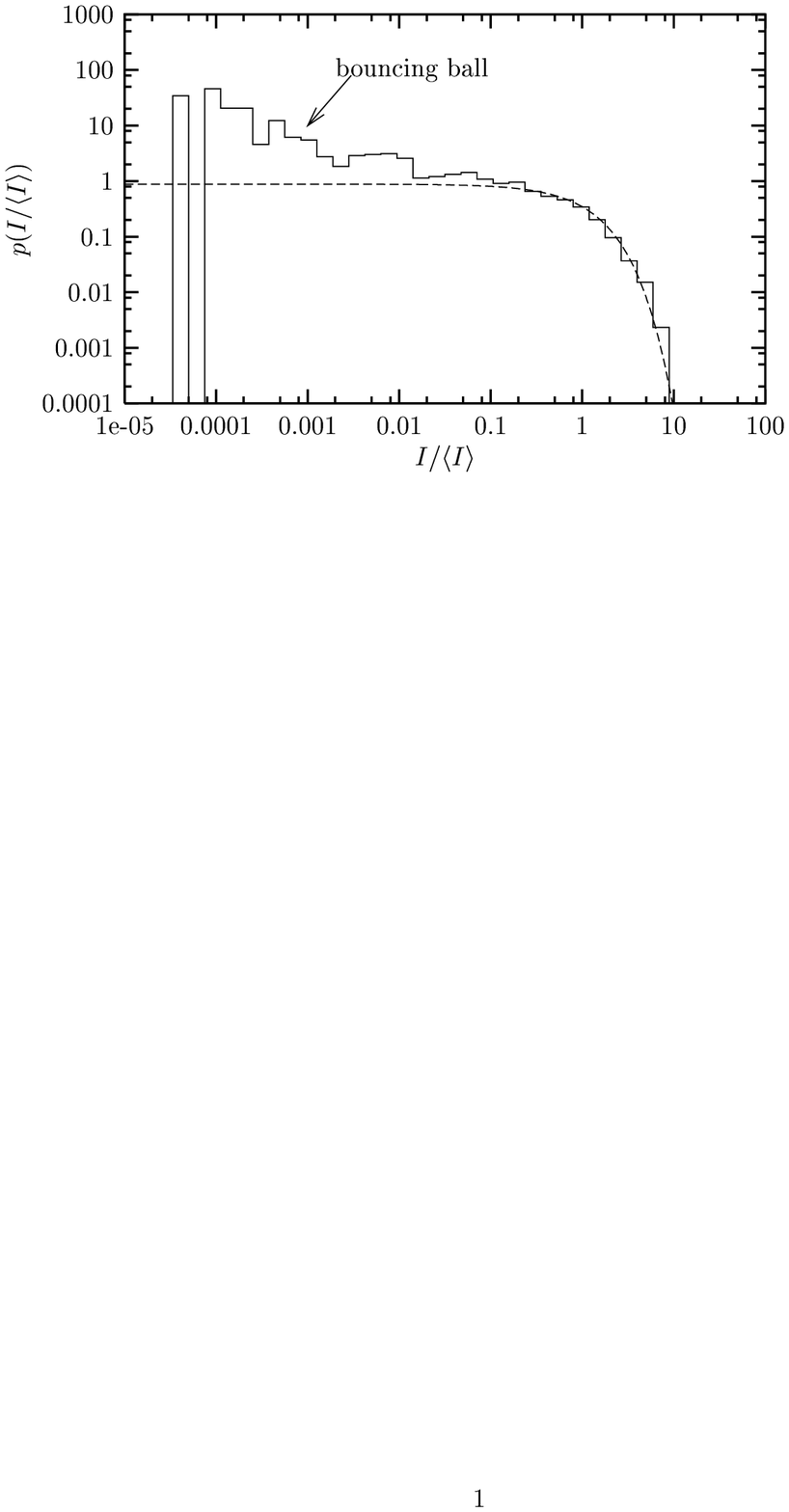,width=3in,bbllx=85pt,bblly=500pt,bburx=445pt,bbury=720pt,clip=}}

\centerline{(b)}

\vskip 0.2in

\caption{Probability distribution of wavefunction
  intensities $I_n = \abs{\protect\braket{q,p}{n}}^2$ for the 1746
  energy levels between $k=100$ and $150$.  (a) For $(q,p)$ situated
  on the bowtie orbit.  Solid line, numerical distribution
  of wavefunction intensities; dashed line, scar-corrected prediction
  of random-matrix theory with one degree of freedom; dotted line,
  random-matrix prediction, uncorrected for scarring. For purposes of
  illustration, the low-intensity
  part is shown in the figure on the left in a semilog plot and the 
  high-intensity tail on the right in a log-log plot. (b) Same as (a)
  but for $(q,p)$ at a generic point in phase space, and the dashed
  line is now the random-matrix theory prediction for two degrees of
  freedom (without scar corrections, but adjusted for the 6\%
  nearly vanishing intensities due to bouncing-ball states).}

\label{fig:intensity-dist}
\end{figure}

\newpage

\begin{figure}

\centerline{
\psfig{file=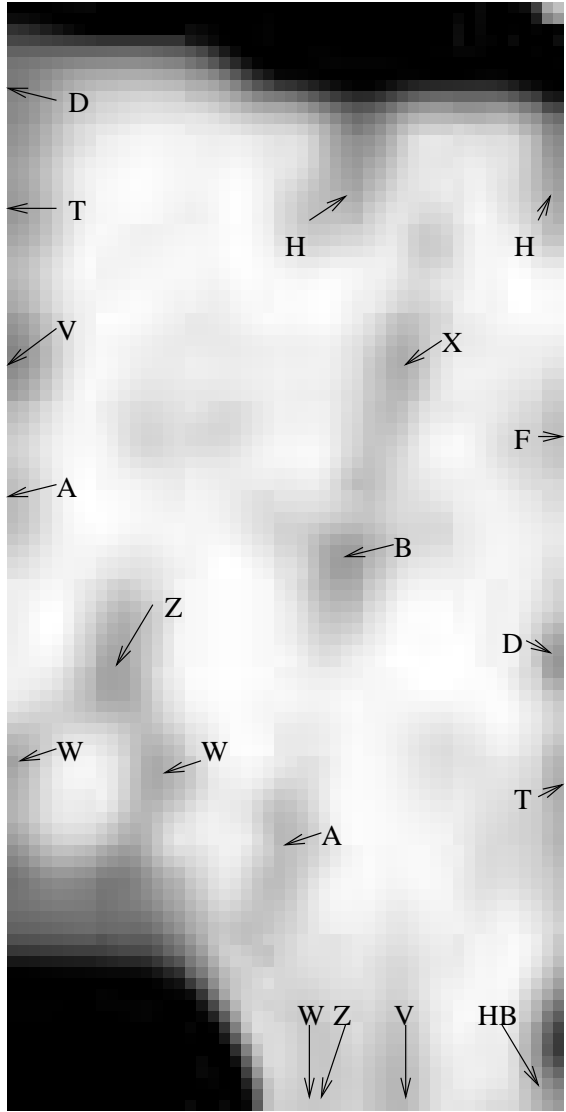,width=3in,bbllx=1pt,bblly=1pt,bburx=216pt,bbury=434pt,clip=}}

\vskip 0.2in

\caption{Principal dynamics-related structures in phase space of the
  inverse-participation ratio plotted in Fig.~\ref{fig:LIPR} and the
  average-return probability plotted in Fig.~\ref{fig:Paa(T)}.  The
  labeling is according to Table~\ref{table:orbits} notation of scars
  on the principal short periodic orbits in Fig.~\ref{fig:Paa(T)}(e),
  in an irreducible one-eighth region of phase space.  The other
  regions of phase space are obtained by symmetry.  
  The most prominent scarred orbits are W, Z, D, B, X and H.}

\label{fig:dynamical-structures}
\end{figure}

\newpage

\begin{figure}

\centerline{
\psfig{file=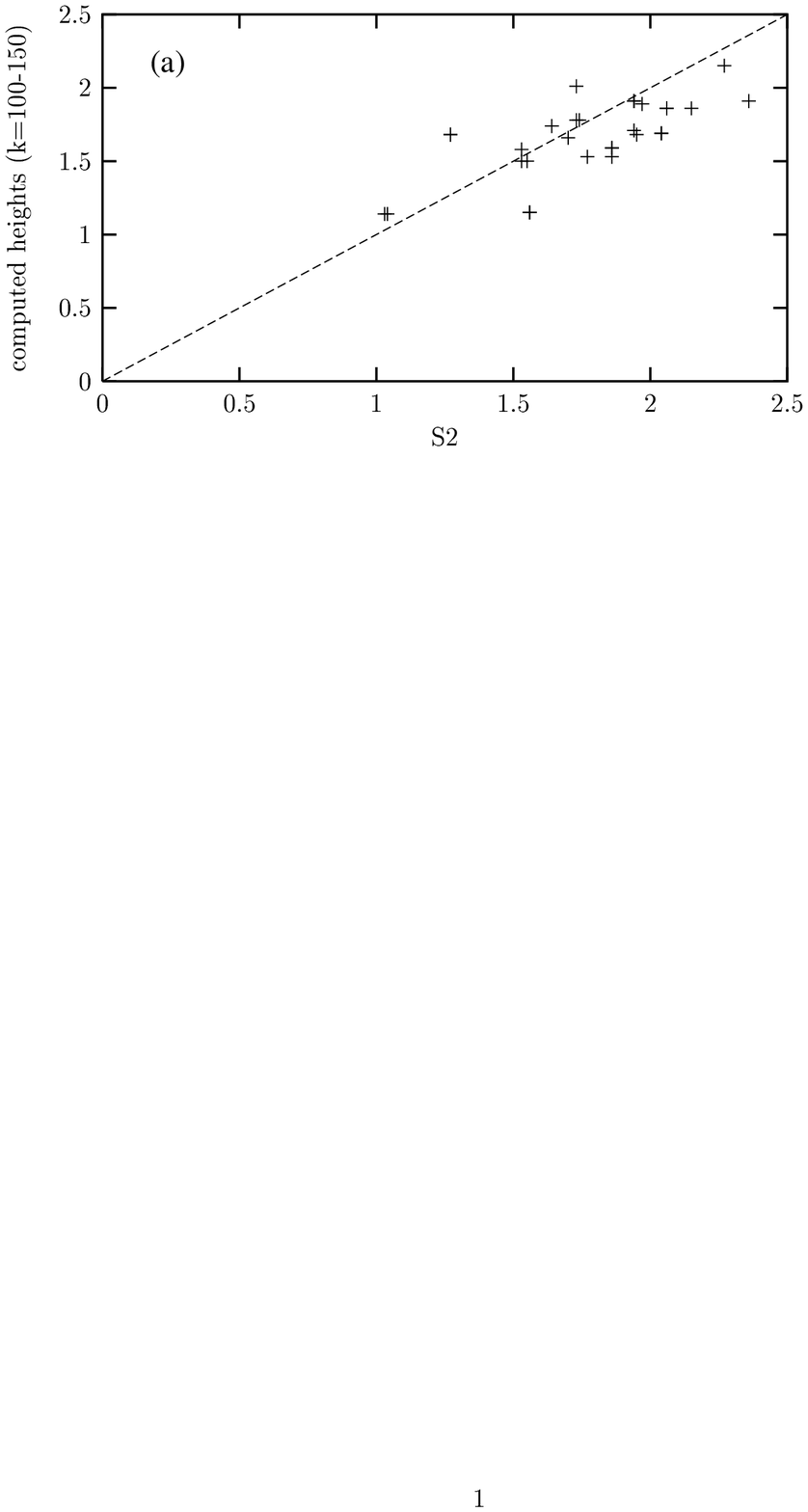,width=3in,bbllx=90pt,bblly=510pt,bburx=455pt,bbury=720pt,clip=}
\psfig{file=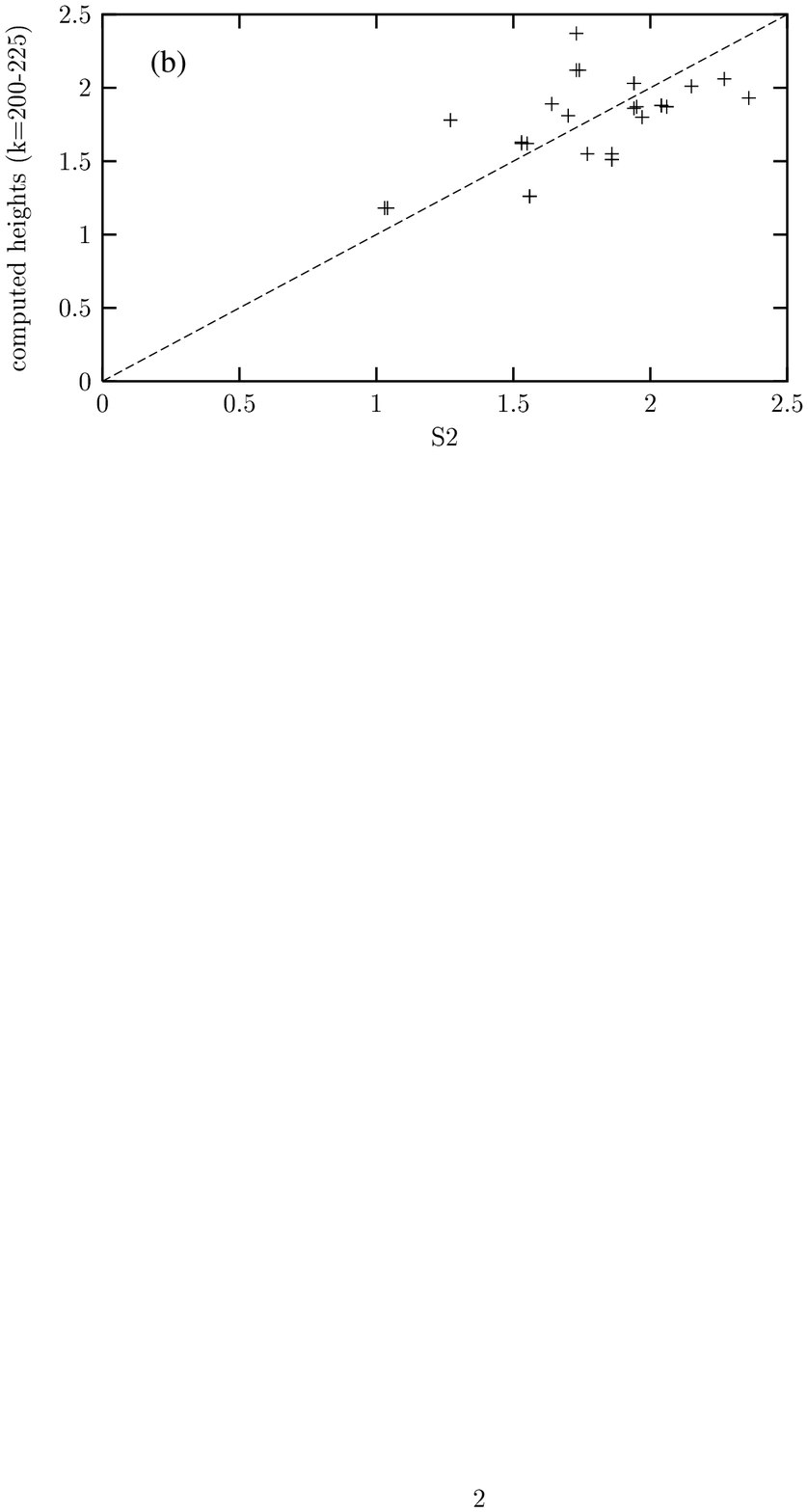,width=3in,bbllx=90pt,bblly=510pt,bburx=455pt,bbury=720pt,clip=}
}

\vskip 0.2in

\centerline{
\psfig{file=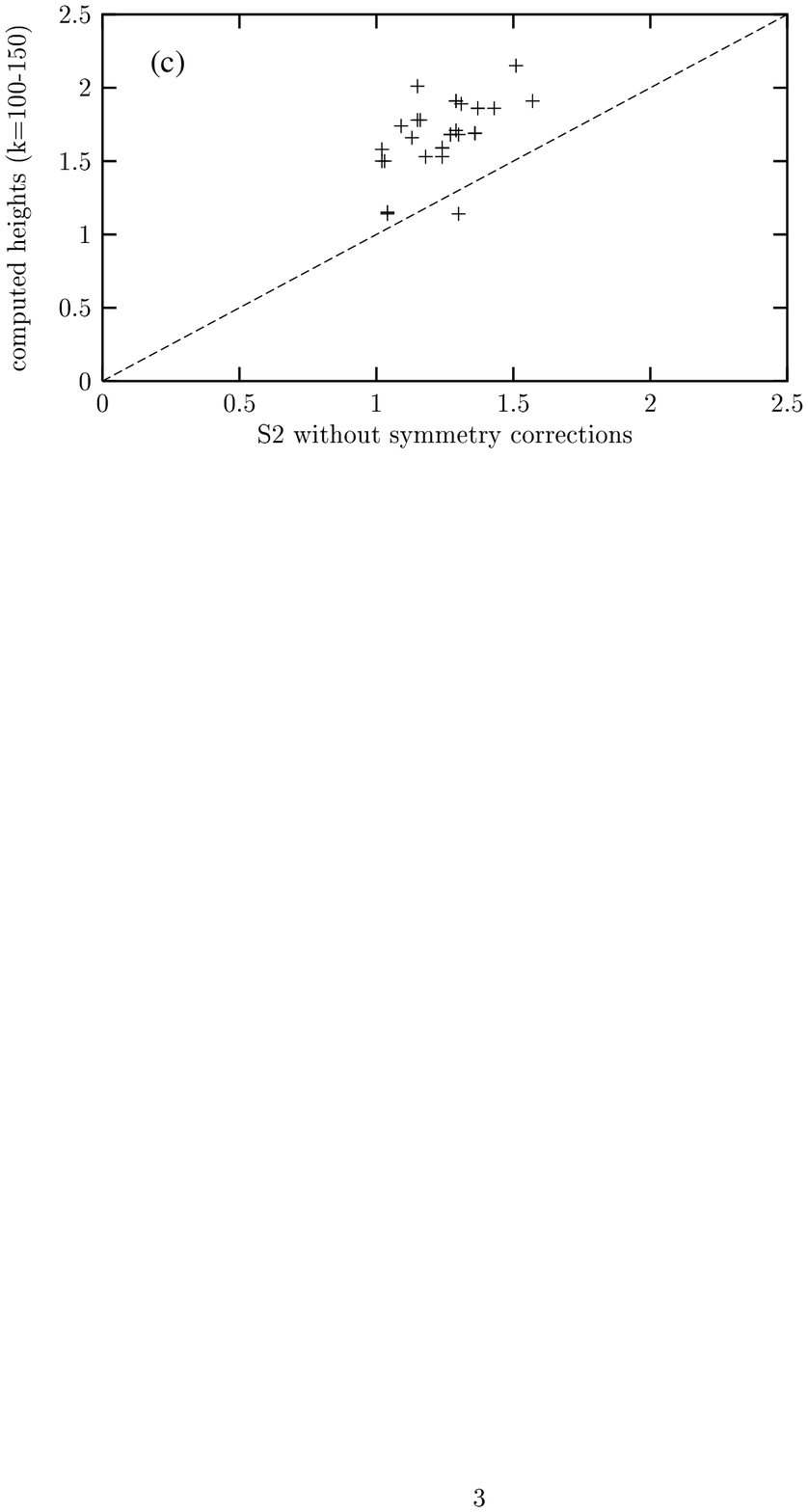,width=3in,bbllx=90pt,bblly=510pt,bburx=455pt,bbury=720pt,clip=}
\psfig{file=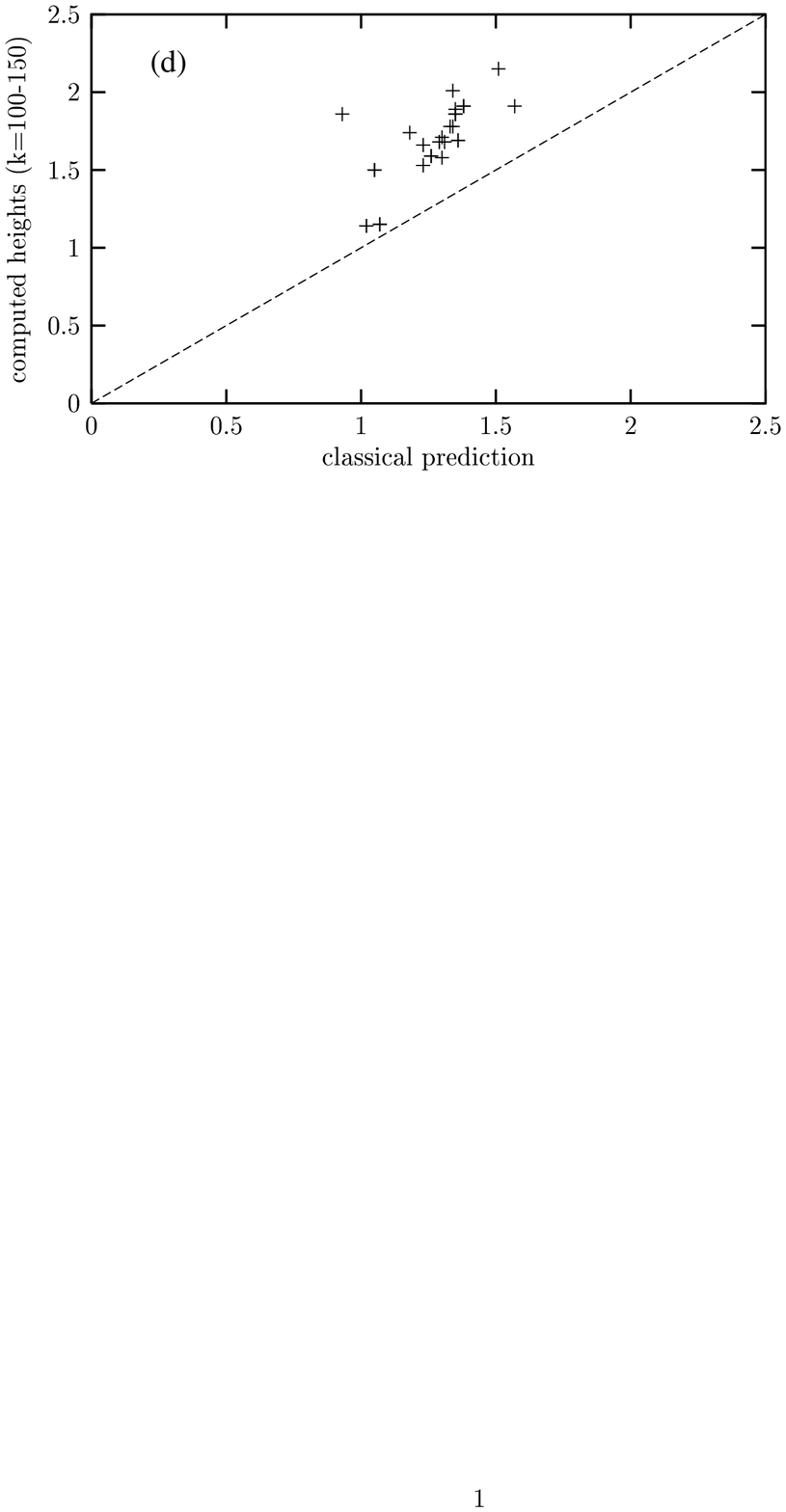,width=3in,bbllx=90pt,bblly=510pt,bburx=455pt,bbury=720pt,clip=}
}

\vskip 0.2in

\centerline{
\psfig{file=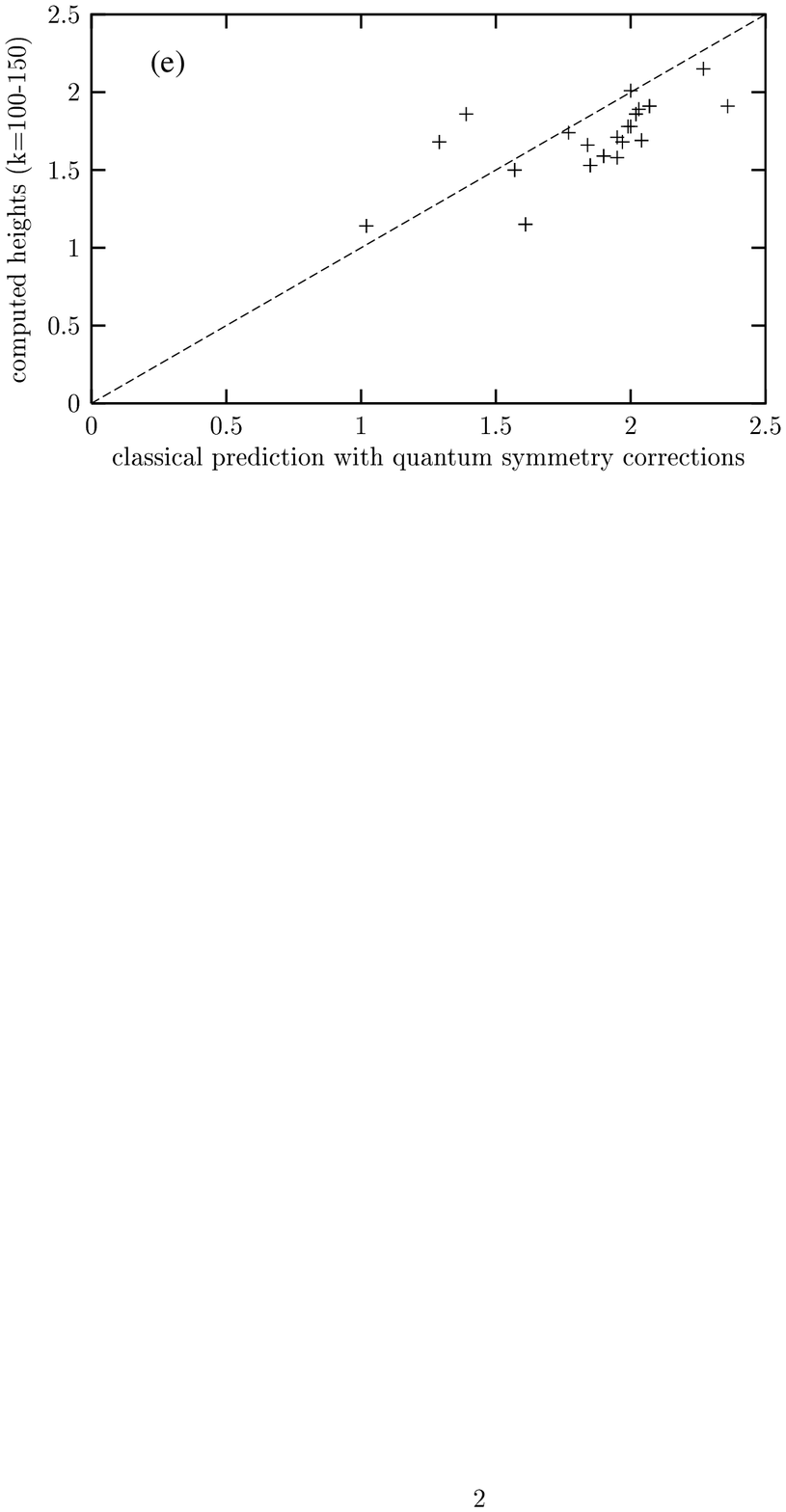,width=3in,bbllx=90pt,bblly=510pt,bburx=455pt,bbury=720pt,clip=}
}

\vskip 0.2in

\caption{Scatter plots of computed peak heights in columns 5 and 6 of
  Table~\ref{table:peak-heights} versus the linearized semiclassical
  predictions $S_2$ in column 4 and the brute-force semiclassical
  prediction in column 7.  (a) Eigenfunctions ranging from $k=100$ to
  $150$, symmetry corrections included in $S_2$; (b) $k=200$ to $225$,
  symmetry corrections again included in $S_2$; (c) $k=100$ to $150$,
  but symmetry corrections {\it not} included in $S_2$.  Without
  symmetry corrections the agreement is much worse.  (d) $k=100$ to
  $150$ versus brute-force classical prediction without symmetry
  correction and (e) with symmetry corrections included.}

\newpage

\label{fig:scatter-plots}
\end{figure}

\newpage

\begin{figure}

\centerline{
\psfig{file=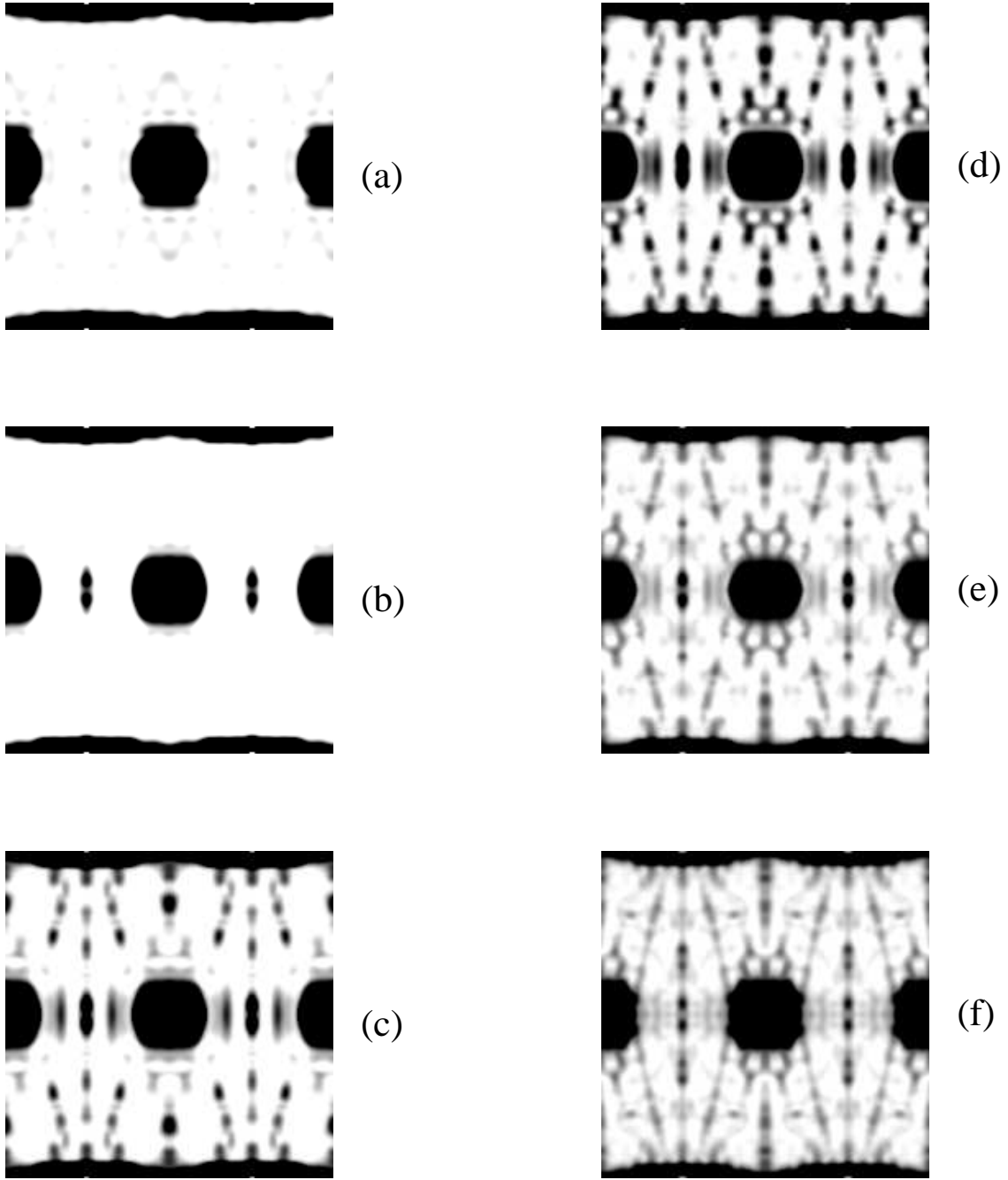,width=6in,bbllx=90pt,bblly=145pt,bburx=520pt,bbury=650pt,clip=}}

\vskip 0.1in
\caption{Quantum-mechanical average-return probability $P_{aa}(T)$
  as defined in Eq.~(\ref{eq:2}), near $k=100$; $q$ and $p$ as before.  (a)
  $T=T_B$, (b) $T=2T_B$, (c) $T=3T_B$, (d) $T=4T_B$, (e) $T=5T_B$ and
  (f) $T=10T_B$.  Note the emergence of periodic orbits at $T=2T_B$ to
  $T=3T_B$.  For $T=10T_B$ the figure is already beginning to resemble
  the LIPR plotted in Fig.~\ref{fig:LIPR}(b).}

\label{fig:Paa(T)}
\end{figure}

\newpage

\begin{figure}

\centerline{
\psfig{file=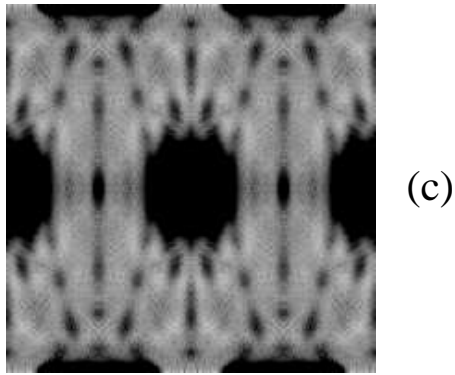,width=2.5in,bbllx=215pt,bblly=145pt,bburx=395pt,bbury=650pt,clip=}}

\vskip 0.1in
\caption{Results of a classical simulation of the return probability
  versus time, $q$ running horizontally from 0 to $4+2\pi$ and $p$
  vertically from $-1$ to 1.  Initial Gaussians were chosen with
  $\sigma^2=0.05$, corresponding to a spread in position space of 0.2
  out of $4+2\pi$, or 2\% of the perimeter.  All of the major periodic
  orbits are present here as short-time recurrences.  (a) $T=5T_B$, (b)
  $T=10T_B$, (c) $T=20T_B$.}

\label{fig:LIPR-classical}
\end{figure}

\end{document}